\documentclass[prd,amsmath,amssymb,twocolumn]{revtex4-1}

\usepackage{graphicx}
\usepackage{color}
\usepackage{amsmath}
\usepackage{amssymb}

\usepackage{bm}
\usepackage{bbm}
\usepackage{MnSymbol}

\newcommand{\CH}{\mathbb{C}\otimes\mathbb{H}}

\newcommand{\CHO}{\mathbb{C}\otimes\mathbb{H}\otimes\mathbb{O}}
\newcommand{\RCHO}{\mathbb{R}\otimes\mathbb{C}\otimes\mathbb{H}\otimes\mathbb{O}}

\newcommand{\CLeight}{\mathbb{C}l(8)}

\newcommand{\CLsix}{\mathbb{C}l(6)}

\newcommand{\CLten}{\mathbb{C}l(10)}

\newcommand{\CLtwon}{\mathbb{C}l(2n)}

\newcommand{\C}{\mathbb{C}}
\newcommand{\R}{\mathbb{R}}

\newcommand{\A}{\mathbb{A}}

\usepackage{xcolor}

\usepackage{color}
 
\begin{document}

\title{An Algebraic Roadmap of Particle Theories  \vspace{2mm}\\ \it Part I:  General construction\rm}

\author{N. Furey}
\affiliation{$ $\\  Iris Adlershof, Humboldt-Universit\"{a}t zu Berlin,\\ Zum Grossen Windkanal 2, Berlin, 12489 \vspace{2mm} \\ furey@physik.hu-berlin.de\\ HU-EP-23/64  }

\begin{abstract} 

Expanding the results of~\citep{AGUTS}, \citep{fh1}, \citep{fh2}, we demonstrate a network of algebraic connections between  \it six well-known particle theories. \rm  These are the Spin(10) model, the Georgi-Glashow model, the Pati-Salam model, the Left-Right Symmetric model, the Standard Model both pre- and post-Higgs mechanism. 

A new inclusion of a quaternionic reflection within the network further differentiates $W^{\pm}$ bosons from the $Z^0$ boson in comparison to the Standard Model.  It may introduce subtle new considerations for the phenomenology of electroweak symmetry breaking.
  \end{abstract}

\color{black}

\maketitle

\section{Introduction}

\begin{center}

 \it You shall know a word by the company it keeps. \rm 

 - J.R. Firth, 1957.

\end{center}

If Quantum Mechanics, \citep{dirac} and Special Relativity, \citep{relativity}, have taught us anything, it is that objects are often best understood not in isolation, but rather, in relation to their peers.  Perhaps the Standard Model of particle physics is no exception. 

In this article, we identify a set of repeating connections, in series and in parallel, between a number of neighbouring particle models.  Among these nine models, six are well-known.  We pinpoint the post-Higgs Standard Model at the cluster's most highly constrained corner.

\begin{figure}[h!]
\begin{center}
\includegraphics[width=10cm]{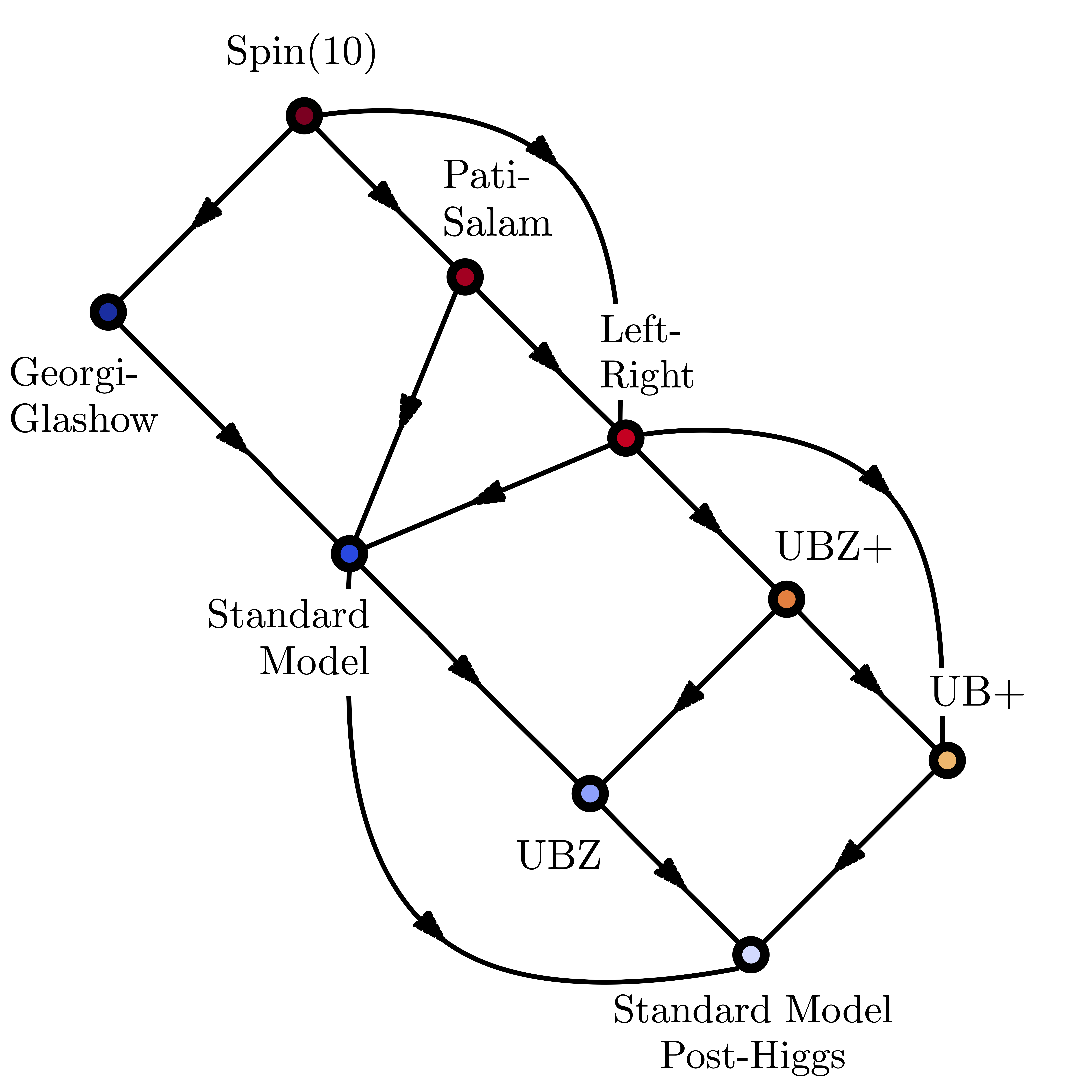}
\caption{\label{words}
Simplified preview of an algebraic particle roadmap.  The detailed version appears in Figure (6) at the end of this article.}
\end{center}
\end{figure}

For a glance ahead, see Figure~(\ref{words}).  There, the network we are about to assemble will be built up using the natural operator spaces of the division algebras $\R, \C, \mathbb{H},$ and $\mathbb{O}.$  These operator spaces (multiplication algebras) are readily described by Clifford algebras.

Since at least as early as the 1970s, physicists and mathematicians have been identifying Clifford algebraic and division algebraic patterns within the architecture of elementary particles.   Now half a century ago, G\"{u}naydin and G\"{u}rsey proposed a quark model based on octonions, \citep{GGquarks}.  Shortly thereafter came a series of papers by Casalbuoni \it et al \rm who proposed particle models based on a large variety of Clifford algebras, \citep{it1}, \citep{it2}, \citep{it3}.  Since these days, many authors have invested a great deal of time and effort to the endeavour.  In what is apologetically far from an exhaustive list, we point out the Clifford algebraic work of Trayling and Baylis (Cl(0,7)) \citep{Greg2001}, Barrett (via NCG) \citep{John2006}, Zenczykowski ($\CLsix$) \citep{Piotr2008}, Connes \it et al \rm (via NCG) \citep{Alain2013},  Stoica ($\CLsix$) \citep{Stoica2017}, Gording and Schmidt-May ($\CLsix$) \citep{Brage2020},  Todorov ($\CLten$) \citep{Ivan2020}, \citep{Ivan2021}, Bor\v{s}tnik (Cl(p,q)) \citep{Norma2023}, the division algebraic work of Conway ($\CH$) \citep{Conway1937}, Silagadze ($J_3(\mathbb{O})$) \citep{Silagadze1994},  Adler ($\mathbb{H}$) \citep{Steve1996}, De Leo ($\mathbb{H}$) \citep{Stefano1996}, Toppan \it et al \rm ($\mathbb{H}, \mathbb{O}$) \citep{Toppan2003}, Baez ($\R, \C, \mathbb{H}$) \citep{Baez2011}, Duff \it et al \rm ($\R, \C, \mathbb{H}, \mathbb{O}$) \citep{Duff2014}, Hughes ($\mathbb{O}$) \citep{mia}, Catto \it et al \rm ($\mathbb{O}$) \citep{Catto2017}, Gresnigt ($\mathbb{O}$) \citep{Niels2018}, Asselmeyer-Maluga ($\mathbb{H}, \mathbb{O}$) \citep{Torsten2019}, Bolokhov ($\mathbb{H}$) \citep{Pasha2019}, Vaibhav and Singh (split-$\mathbb{H}$ and split-$\mathbb{O}$) \citep{Tejinder2021}, Boyle ($J_3(\mathbb{O})$) \citep{boyle1}, Jackson ($\mathbb{O}$) \citep{David2021}, Hunt ($\R, \C, \mathbb{H}, \mathbb{O}$) \citep{Bruce2022}, Lasenby ($\mathbb{O}$) \citep{Anthony2023}, Manogue \it et al \rm ($E_8$) \citep{Corinne2022}, \citep{ms}, Hun Jang, (Hypercomplex) \citep{Hun2023}, Hiley (split-$\mathbb{H}$) \citep{Basil2023},  and most recently Penrose (split-$\mathbb{O}$) \citep{Roger2023},  the $\RCHO$ work of Dixon, \citep{Dixon1999}, \citep{Dixon_recent}, Castro Perelman, \citep{Carlos2019}, Chester \it et al, \rm \citep{David2023}, K\"{o}plinger, \citep{Jens2023}.

In earlier years, several authors have employed the division algebras in order to break a variety of symmetries.  In \citep{GGquarks}, G\"{u}naydin and G\"{u}rsey made use of an octonionic imaginary unit so as to break $G_2\mapsto $ SU(3) in the context of a quark model.  In~\citep{Dixon1999}, Dixon made use of two octonionic projection operators in order to reduce Spin(1,9)$\times$SU(2) to Spin(1,3) and a non-chiral representation of the Standard Model gauge group.  These groups acted on a fermionic space described by two copies of $\RCHO,$ later to be known as the Dixon algebra.  In~\citep{boyle1},  Boyle studied an $E_6$ model in the context of a complexified version of the exceptional Jordan algebra, $J_3(\mathbb{O})$.  There, he found that a single octonionic imaginary unit may break Spin(10)$\subset E_6$ to the Left-Right symmetric model.  

Other related symmetry breaking steps were described in the 1970s by Casalbuoni and Gatto in a footnote of~\citep{it3}.  There a Spin(2n) $\mapsto$ SU(n) breaking is effected by requiring the invariance of a fermionic monomial.  Similarly, in \citep{Greg2001}, Trayling and Baylis  proposed fixing the sterile neutrino in the context of Cl(0,7).  Subsequently Todorov also proposed fixing the sterile neutrino in \citep{Ivan2020} in the context of $\CLten$.  These symmetry breaking steps are closely tied to Baez and Huerta's proposal to fix a fermionic volume element in~\citep{AGUTS}.

This paper is the first of a series; see also  \citep{fr2}, \citep{fr3}.  Although the model described here was non-trivial to find, readers may appreciate that a large number of its results can be confirmed easily.  

\section{In context}

The particle roadmap introduced in this article extends directly from previous findings of Baez and Huerta~\citep{AGUTS}, and Furey and Hughes~\citep{fh1}, \citep{fh2}.  We summarize these earlier findings here.  This section provides the historical background for this article, but is not strictly necessary to understand its results.

\subsection{Algebra of grand unified theories\label{aguts}}   

It is a surprisingly little-known fact amongst particle physicists that many of our most well-studied theories are interrelated.

The birth of the Spin(10) model in the mid-1970s came only a few hours before that of Georgi and Glashow's SU(5) model~\citep{so10hist}, \citep{su5}.  On its own, Georgi and Glashow's model posits a seemingly \it ad hoc \rm fermionic particle content as the $\mathbf{10}\oplus \mathbf{5^*}$ of SU(5).  Why, one might wonder, this curious combination of irreducible representations?  

However, embedding these irreps inside the proposed $\mathbf{16}$ of Spin(10)  justifies the representation structure, while adding in a sterile neutrino in the form of an $SU(5)$ singlet. 

\begin{equation} \begin{array}{ccc}
\mathfrak{so}(10) &\supset&  \mathfrak{su}(5) \vspace{2mm}\\
\mathbf{16} &\hspace{1cm}\mapsto\hspace{1cm}& \mathbf{10}\oplus \mathbf{5^*}\oplus \mathbf{1}
\end{array}\end{equation}

\noindent Hence the Spin(10) model offers guidance for its younger SU(5) sibling.

In the same era, another (partially) unified theory was constructed by Pati and Salam,~\citep{PS}.  Their intention was to capitalize on  an observed pattern-matching between quarks and leptons.  Via the group SU(4)$\times$SU(2)$\times$SU(2), the Standard Model's three ``red, green, blue" quarks colours were augmented to include a fourth ``lilac" lepton colour.  Furthermore, a symmetry between left- and right-handed particles was conjectured.  Curiously enough, Pati and Salam's model also fits neatly into the $\mathbf{16}$ of Spin(10):

\begin{equation} \begin{array}{ccc}
\mathfrak{so}(10) &\supset&  \mathfrak{su}(4)\oplus  \mathfrak{su}(2) \oplus \mathfrak{su}(2)\vspace{2mm}\\
\mathbf{16} &\hspace{1cm}\mapsto\hspace{1cm}& \left(\mathbf{4}, \mathbf{2}, \mathbf{1} \right)\oplus \left(\mathbf{4^*}, \mathbf{1}, \mathbf{2} \right)
\end{array}\end{equation}

\noindent Again, we find that the Spin(10) model justifies for the Pati-Salam model an otherwise arbitrary choice in fermion representations.

Now, more surprising than the kinship between the Spin(10) and SU(5) models, or the kinship between Spin(10) and Pati-Salam models, is an unexpected kinship between SU(5), Pati-Salam, and the Standard Model.  That is, in~\citep{AGUTS}, Baez and Huerta report that the Standard Model's gauge group coincides exactly with the intersection between SU(5) and Pati-Salam symmetries.  The following Figure~(\ref{jj}) relates these four well-studied particle models.

\begin{figure}[h!]
\begin{center}
\includegraphics[width=7.5cm]{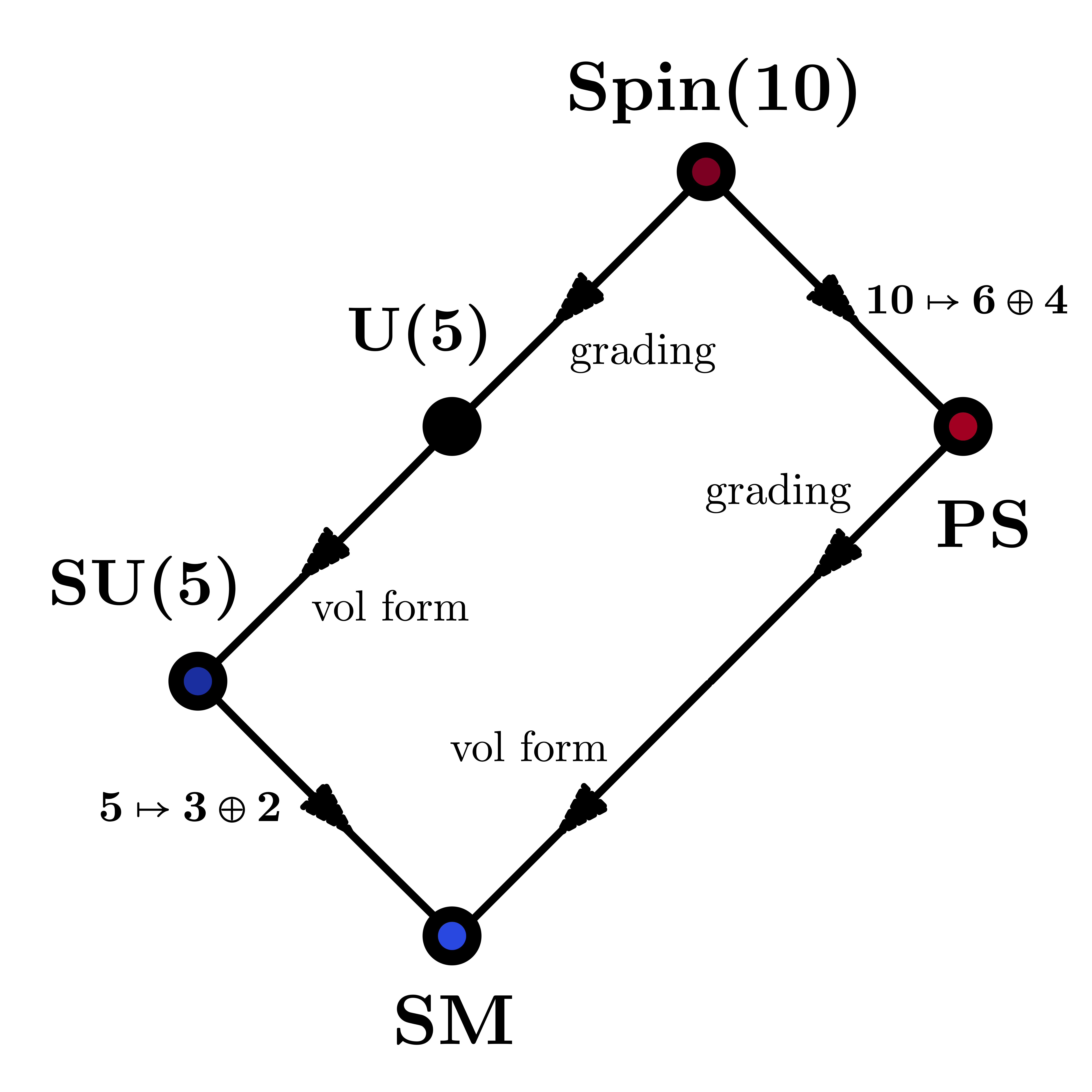}
\caption{\label{jj}
Baez and Huerta, \citep{AGUTS}, explain how Standard Model symmetries (SM) result from the intersection of Georgi and Glashow's SU(5) and the Pati-Salam (PS) symmetries SU(4)$\times$SU(2)$\times$SU(2).  Spin(10) may be seen to break to PS by preserving the splitting $\mathbf{10}\mapsto \mathbf{6}\oplus \mathbf{4}$ of SO(10)'s defining representation.  Alternatively, Spin(10) may be seen to break to SU(5) by preserving a grading and volume form of the spinor space $\Lambda \C^5$. }
\end{center}
\end{figure}

But what could prompt the symmetries to break in this way?  Reference~\citep{AGUTS} lists three independent conditions:

\vspace{2mm}

1.  A requirement that the defining $\mathbf{10}$ of $\mathfrak{so}$(10) splits into $\mathbf{6}\oplus \mathbf{4}$, (this breaks $\mathfrak{so}(10) \mapsto  \mathfrak{su}(4)\oplus  \mathfrak{su}(2) \oplus \mathfrak{su}(2)$),

2.  The preservation of a $\mathbb{Z}$-grading on the spinorial $\Lambda \C^5 \simeq \mathbf{16}\oplus \mathbf{16^*}$ of $\mathfrak{so}$(10), (this breaks $\mathfrak{so}(10) \mapsto  \mathfrak{u}(5)$),

3.  The invariance of a volume form in the spinorial $\Lambda \C^5 \simeq \mathbf{16}\oplus \mathbf{16^*}$ of $\mathfrak{so}$(10), (this breaks $\mathfrak{u}(5) \mapsto  \mathfrak{su}(5)$).

\vspace{2mm}

\noindent Here, $\Lambda \C^5$ denotes the exterior algebra generated by vectors of 5 complex dimensions.  See also footnotes in~\citep{it3}.

From where do these conditions arise?  Baez and Huerta  encourage readers to decipher  the meaning behind these mysterious constraints.

\subsection{Division algebraic symmetry breaking}

That the first constraint, 
\begin{equation} \begin{array}{ccc}
\mathfrak{so}(10) &\mapsto&  \mathfrak{so}(6)\oplus  \mathfrak{so}(4) \vspace{2mm}\\
\mathbf{10} &\hspace{1cm}\mapsto\hspace{1cm}& \mathbf{6}\oplus\mathbf{4}
\end{array}\end{equation}
\noindent results from preserving an octonionic volume element was first discovered in~\citep{fh2}.  
\it This octonionic structure may be seen to be responsible for sending the Spin(10) model to the Pati-Salam model, and the SU(5) model to the Standard Model. \rm

\begin{figure}[h!]
\begin{center}
\includegraphics[width=9.3cm]{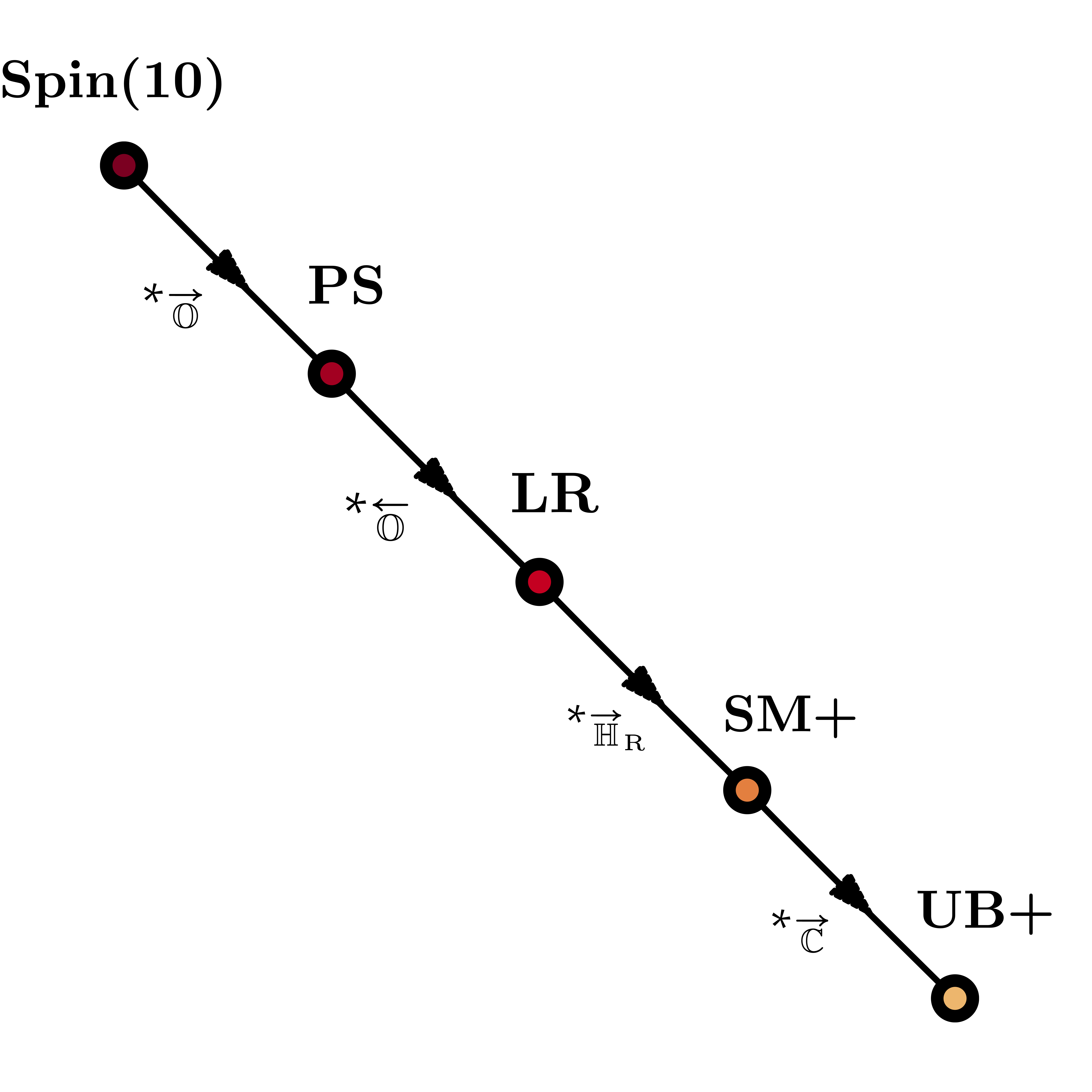}
\caption{\label{casc}
References \citep{fh1}, \citep{fh2} introduced a cascade of breaking symmetries. The Spin(10) model breaks to the Pati-Salam model, (PS), via the octonions.  The Pati-Salam model breaks to the Left-Right Symmetric model, (LR), again via the octonions.   The Left-Right Symmetric model breaks to the Standard model + B-L, (SM+), via the quaternions.  The Standard model + B-L breaks to the Standard Model's unbroken symmetries, together with B-L, (UB+), via the complex numbers. Two open questions remained in this model:  (1) Why the lingering B-L symmetry?  (2) Could there be a more natural way to introduce chirality in the quaternionic ``$\mathbb{H}_{\textup{R}}$" step described in~\citep{fh2}?}
\end{center}
\end{figure}

As will be explained in detail later on in this text, the Spin(10) $\mapsto$ Pati-Salam and SU(5) $\mapsto$ Standard Model transitions occur upon the requirement that the symmetries be invariant under a certain type of \it octonionic reflection. \rm  Please see Figure~(\ref{casc}).  Going beyond, it was found in \citep{fh1}, \citep{fh2}  that invariance under a complementary octonionic reflection furthermore sends this Pati-Salam model ($\mathfrak{so}(6)\oplus  \mathfrak{so}(4) =  \mathfrak{su}(4)\oplus  \mathfrak{su}(2) \oplus \mathfrak{su}(2)$) to the  Left-Right Symmetric model ($\mathfrak{su}(3)\oplus  \mathfrak{su}(2) \oplus \mathfrak{su}(2)\oplus \mathfrak{u}(1)$).  Then, invariance under a quaternionic reflection sends the Left-Right Symmetric model to the Standard Model, augmented with a B-L symmetry ($\mathfrak{su}(3)\oplus  \mathfrak{su}(2) \oplus \mathfrak{u}(1)\oplus \mathfrak{u}(1)$).  Finally, \citep{fh1},  invariance under a complex conjugate reflection sends the pre-Higgs Standard Model  ($\mathfrak{su}(3)\oplus  \mathfrak{su}(2) \oplus \mathfrak{u}(1)$) to the post-Higgs Standard Model  ($\mathfrak{su}(3)\oplus\mathfrak{u}(1)$).  

In short, invariance under $\mathbb{O},$ $\mathbb{H},$ and $\C$ reflections links five well-established particle models in a cascade of breaking symmetries.

However, as with~\citep{AGUTS}, some open questions remain.  Namely, in the quaternionic Left-Right Symmetric $\mapsto$ Standard Model step, 

1.  An additional  B-L symmetry persisted,

2.  Chirality was introduced by hand.

\noindent Could it be  possible to tame these unruly features?

In this article, both of these issues are addressed simultaneously.  Furthermore,  a concise answer s offered to those challenges posed by Baez and Huerta as to where from their mysterious conditions may be seen to originate.  

As will be seen, the overarching theme may be described as the implementation of \it simultaneous group actions \rm on the fermionic representations.  These simultaneous group actions, or more precisely, Lie algebra actions, allow one to upgrade and merge the diagrams of Figures~\ref{jj} and \ref{casc}.  The result (Figure~\ref{5step}) is a detailed roadmap of particle models. 

In order to keep the discussion as concise as possible, we will work on the level of Lie algebras.  Those interested in  group theoretic details are encouraged to consult~\citep{AGUTS}.

\subsection{A consistency condition}

It is hard not to notice that, when augmented by a sterile neutrino, the fermionic representations of the Standard Model materialize as broken pieces of the $\mathbf{16}$ spinor representation of Spin(10).  However, the $\mathfrak{su}(3)\oplus  \mathfrak{su}(2) \oplus \mathfrak{u}(1)$ Standard Model symmetries acting on this representation space are unmistakably \it unitary. \rm  Could there be more than one type of symmetry constraint working in concert?  

On a not-obviously related front, let us point out the \it minimal left ideal \rm construction of spinors.  As will be explained in detail shortly, such spinors may be constructed  from the $2n$ generating $\Gamma$-matrices of certain $\C l(2n)$ Clifford algebras.  Collectively, linear combinations of these $\Gamma$-matrices form the defining $2n$ vector representation of the Clifford algebra's $\mathfrak{so}$(2n) symmetries.  But this then begs the question:  \it If the spinor is constructed directly from these vectors, should it not also transform under a group action induced by them? \rm

In this article, we entertain the idea that Standard Model fermions should simultaneously transform under both the known $\mathfrak{spin}(2n)$ spinor action, and also under the known $\mathfrak{so}(2n)$ multivector action.  This consistency condition reduces $\mathfrak{spin}(2n)$ to $\mathfrak{su}(n)$.   In one stroke, this single requirement addresses both the chirality and B-L issues plaguing~\citep{fh2}.   It also fuses Baez and Huerta's conditions 2. and 3. in Section~\ref{aguts} into a single constraint. Here, we derive the result  as a consistency condition in the construction of fermions as multivectors.

The main results of these papers were first made public in an abstract  circulated for the Perimeter Institute Octonions and Standard Model conference in 2021.  We include the abstract here:~\citep{abstract}.

\subsection{Triality's example}

As an aside, we mention that the idea of simultaneous group actions is not new to representation theory, \citep{mia}, \citep{ms}.  A phenomenon known as \it octonionic triality \rm has three copies of $\mathbb{O}$ each transforming under Spin(8).  Distinct $\mathfrak{so}$(8) actions on these 8D spaces allow one copy to be identified with the $\psi$ spinor, another with the $\widetilde{\psi}$ conjugate spinor, and the final copy with the vector representation, $V$.  Now, introducing a requirement that each of the corresponding group actions coincide then reduces the original $\mathfrak{so}$(8) symmetry to $\mathfrak{g}_2$.  While $\mathfrak{so}$(8) generated the triality symmetry of the octonions, $\mathfrak{g}_2$ generates their automorphisms.  

Parallels between automorphisms embedded inside triality symmetries, and the standard model internal symmetries embedded inside $\mathfrak{spin}(10)$ will be made in the third paper of this series,~\citep{fr3}.

\section{Simultaneous group actions of the multivector type}

In this section, we offer an explanation for the breaking of symmetry from $\mathfrak{spin}(2n) \mapsto \mathfrak{su}(n)$, first proposed in~\citep{abstract}.  In known models, this corresponds to transitions 
$$ \begin{array}{rcl} \textup{Spin(10)} &\mapsto&  \textup{SU(5),} \vspace{2mm}\\

 \textup{Pati-Salam} &\mapsto&  \textup{Standard Model,}  \vspace{2mm}\\

 \textup{Left-Right Symmetric} &\mapsto&  \textup{Standard Model.}
\end{array}$$
\noindent While algebraic \it single-step \rm methods breaking $\mathfrak{spin}(2n) \mapsto \mathfrak{u}(n)$ are ubiquitous in the literature, algebraic single-step methods breaking $\mathfrak{spin}(2n) \mapsto \mathfrak{su}(n)$ are not.

\subsection{Three types of spinor}

We argue in this subsection that not all spinor constructions are created equal.  We compare three inequivalent constructs with increasing complexity. 

(I)  The most common description of a spinor in the literature is simply as a column vector (equivalently, single-indexed tensor) with entries in $\mathbb{R}$ or $\C$.

(II)  As an alternative to column vectors, certain spinors may be constructed by upgrading these vector spaces to exterior algebras, $\Lambda \C^{n}$.  In this case, the spinor is endowed with a $\mathbb{Z}$-grading and a concept of a wedge product on its elements.  

(Aside) It has long been known that this construction leads  to a complementary description in terms of differential forms and vector fields, eg~\citep{it3}.  Namely,
\begin{equation}  a_j = \textup{d} x^j  \wedge  \hspace{1cm} a_j^{\dagger} = i\frac{\partial}{\partial x^j},
\end{equation}
\noindent where the $\{a_j\}$ represent the $n$ generating elements of $\Lambda \C^{n}$.

Augmenting spinors-as-column-vectors to spinors-as-exterior-algebras introduces a structural richness that would otherwise be invisible in the more common column vector approach.  However, this is not the end of the line.

(III) The construction of spinors as \it minimal left ideals \rm (MLI) takes this $\Lambda \C^n$, and  adds another layer of structure beyond it. As will be described shortly, this extra structure appears in the form of a non-trivial vacuum state.  The MLI spinor construction leads to a more unified description by embedding the spinor directly into the Clifford algebra.  We now introduce this construction to readers not already familiar. 

\subsection{Minimal left ideal construction \label{MLIconstr}}

\vspace{2mm}

Consider a Clifford algebra, $\CLtwon$, over the complex numbers, and generated by basis vectors $\Gamma_k$ for $k\in\{1,\dots, 2n\}$, and $n\in\mathbb{N}>0$.  Let $\dagger$ denote an anti-linear involution such that  $(\Gamma_{k_1}\Gamma_{k_2})^{\dagger} = \Gamma_{k_2}^{\dagger}\Gamma_{k_1}^{\dagger}$.  In this setup, we are free to choose $\Gamma_k$ such that $\Gamma_k^{\dagger} = -\Gamma_k,$ and

\begin{equation}\label{CA} \{ \Gamma_{k_1}, \Gamma_{k_2}\} := \Gamma_{k_1}\Gamma_{k_2} +  \Gamma_{k_2}\Gamma_{k_1} = -2\delta_{k_1k_2},
\end{equation}

\noindent and we will indeed do so. 

From here, we  construct an alternative generating basis for $\CLtwon$ as consisting of $n$ raising operators, $a_j$, and $n$ lowering operators, $a^{\dagger}_j$, for $j\in \{1,2,\dots n\}$.  (Naming conventions were chosen so that this work matches previous articles, \citep{fh1}, \citep{fh2}.)  Explicitly, 

\begin{equation} \label{a2g}
a_j := \frac{1}{2} \left( -\Gamma_j + i \Gamma_{n+j} \right) \hspace{1cm} 
a^{\dagger}_j := \frac{1}{2} \left( \Gamma_j + i \Gamma_{n+j} \right).
\end{equation}
\noindent From these definitions and equations~(\ref{CA}), we have that 
\begin{equation}\begin{array}{l}
\{a_{j_1}, a_{j_2} \} =\{a_{j_1}^{\dagger}, a_{j_2}^{\dagger} \} = 0 \vspace{2mm}\\
\{a_{j_1}, a^{\dagger}_{j_2} \} =\delta_{j_1j_2}
\end{array}\end{equation}
\noindent for $j_1, j_2 \in \{1,2,\dots n\}$.  We define $\Omega := a_1a_2 \cdots a_n$.  Then we form a hermitian \it vacuum state \rm  as 
\begin{equation}\label{vac} v := \Omega^{\dagger}\Omega = a_n^{\dagger} \cdots a_2^{\dagger} a_1^{\dagger}a_1 a_2 \cdots a_n .
\end{equation}
\noindent We are now ready to define a minimal left ideal, $\Psi,$ as
\begin{equation}  \Psi := \CLtwon \hspace{.5mm}v.
\end{equation}
\noindent These raising and lowering operators allow us to make apparent a $\mathbb{Z}$-grading within $\Psi$.  It can be shown that 
\begin{equation}\begin{array}{lc}  \label{tower}
\Psi = & z_{12\dots n} \hspace{.5mm}a_1 a_2 \cdots a_n v \vspace{2mm}\\
& \vdots \vspace{2mm}\\
&+\hspace{1mm} z_{12}\hspace{.5mm} a_1a_2v + z_{13}\hspace{.5mm} a_1a_3v  \dots + z_{n-1\hspace{.5mm} n} \hspace{.5mm}a_{n-1}a_nv\vspace{2mm}\\
&+\hspace{1mm} z_1\hspace{.5mm} a_1v + z_2 \hspace{.5mm}a_2v  \dots + z_n \hspace{.5mm}a_nv\vspace{2mm}\\
&+ \hspace{1mm}z_0 \hspace{.5mm}v
\end{array}\end{equation}
\noindent for $z_0, z_j, \dots, z_{12\dots n} \in \C$.  From here, it is straightforward to see that $\Psi$ takes the form of a Fock space, familiar to physicists.   

It can be shown that the omission of $v$ in equation~(\ref{tower}) leads to an exterior algebra.  Hence, the MLI construction may be seen to bestow upon an exterior algebra the extra structure of a non-trivial vacuum state.

\subsection{Adjoint, vector, and spinor representations}

It is known, \citep{FH}, that the  Lie algebra $\mathfrak{so}(2n)$ may be represented by bivectors $r_{k_1k_2} \Gamma_{k_1}\Gamma_{k_2}$ for $r_{k_1k_2}\in\mathbb{R}$.  Here, $k_1, k_2 \in \{1,2,\dots 2n\}$, and $k_1\neq k_2$. Multiplication is given by the commutator, 
$$[ r_{k_1k_2} \Gamma_{k_1}\Gamma_{k_2}, \hspace{1mm}r_{k_3k_4} \Gamma_{k_3}\Gamma_{k_4}].$$

Similarly, the defining $2n$-dimensional vector representation of $\mathfrak{so}(2n)$ may be written as $V := \sum_{k=1}^{2n}V_k \hspace{.5mm}\Gamma_k$ for $V_k\in\mathbb{R}$.  Its infinitesimal transformations under $\mathfrak{so}(2n)$ are then given by 
\begin{equation}\label{vector}  \delta V = [ r_{k_1k_2} \Gamma_{k_1}\Gamma_{k_2}, \hspace{.5mm}V].
\end{equation}
\noindent It is important to note that in this formalism, basis vectors are viewed as carrying the transformations, not the coefficients.  

Finally, infinitesimal transformations on the $2^n$ $\C$-dimensional minimal left ideal $\Psi$ is given by 
\begin{equation}\label{dPsi}  \delta \Psi = r_{k_1k_2} \Gamma_{k_1}\Gamma_{k_2}\Psi.
\end{equation}

\subsection{$\mathfrak{spin}(2n)$ $\mapsto$ $\mathfrak{su}(n)$ via the \\ multivector condition}

In equation~(\ref{tower}), we explained that spinors can be constructed purely from the generating  $\{a_1, a_2, \dots a_1^{\dagger}, a_2^{\dagger},\dots\}$ of the Clifford algebra.  Using equation~(\ref{a2g}), we may then express this spinor purely in terms of the $\{\Gamma_k\}$.  Explicitly, these minimal left ideals are composed of multivectors
\begin{equation} \Psi = c_0 + c_{i_1} \Gamma_{i_1} + c_{i_2i_3} \Gamma_{i_2}\Gamma_{i_3} + \dots,
\end{equation}
\noindent for $i_1,i_2,i_3,\dots\in \{1,2,\dots 2n\}$ and for some $c_0, c_{i_1}, c_{i_2i_3},\dots$ $\in\C$.  

As mentioned in equation~(\ref{vector}), these $\{\Gamma_k\}$ form the vector representation $V := \sum_{k=1}^{2n}V_k \hspace{.5mm}\Gamma_k$ of $\mathfrak{so}(2n)$,
\begin{equation}  \label{vect} \delta V = [ r_{k_1k_2} \Gamma_{k_1}\Gamma_{k_2}, \hspace{.5mm}V].
\end{equation}
\it From this vantage point, it is then natural to wonder:  Shouldn't the generating vectors  induce a transformation rule on the minimal left ideals that were built from them?  \rm

Equation (\ref{vect}) supplies a derivation.  One would  expect an induced infinitesimal transformation of $\Psi$ to materialize as
\begin{equation}\begin{array}{l}\label{der}  \delta \Psi \vspace{2mm}\\= \delta c_0 +  c_{i_1} \hspace{.5mm}\delta \Gamma_{i_1} + c_{i_2i_3}\hspace{.5mm}\delta (\Gamma_{i_2}) \Gamma_{i_3} + c_{i_2i_3}\Gamma_{i_2} \hspace{.5mm}\delta (\Gamma_{i_3}) + \dots \vspace{2mm}\\
 = c_{i_1} \hspace{.5mm}[ r_{k_1k_2} \Gamma_{k_1}\Gamma_{k_2}, \hspace{1mm} \Gamma_{i_1} ]+ c_{i_2i_3}\hspace{.5mm}[ r_{k_1k_2} \Gamma_{k_1}\Gamma_{k_2}, \hspace{1mm} \Gamma_{i_2}] \hspace{.5mm}\Gamma_{i_3} \vspace{2mm}\\
\hspace{2.5mm}+\hspace{1mm} c_{i_2i_3}\Gamma_{i_2} \hspace{.5mm}[ r_{k_1k_2} \Gamma_{k_1}\Gamma_{k_2}, \hspace{1mm} \Gamma_{i_3}]+  \dots ,
\end{array}\end{equation}
\noindent which simplifies to 
\begin{equation}\label{v}  \delta \Psi = [ r_{k_1k_2} \Gamma_{k_1}\Gamma_{k_2}, \hspace{.5mm}\Psi]\hspace{1mm}.
\end{equation}
\noindent However, we already defined the infinitesimal transformation of spinors in equation~(\ref{dPsi}) as 
\begin{equation} \label{s} \delta \Psi = r_{k_1k_2} \Gamma_{k_1}\Gamma_{k_2}\Psi.
\end{equation}
\noindent So, which shall it be:  equation (\ref{v}) or equation (\ref{s})?

Let us entertain the idea that \it both \rm symmetry actions~(\ref{v}) and (\ref{s}) be simultaneously obeyed.  We will refer to this consistency condition as the \it multivector condition. \rm 

Readers may confirm that only an $\mathfrak{su}(n)$ subalgebra of $\mathfrak{so}(2n)$ survives the multivector condition.
\begin{equation} \begin{array}{c}\label{svconstraint} \delta \Psi = [ r_{k_1k_2} \Gamma_{k_1}\Gamma_{k_2}, \hspace{.5mm}\Psi]\hspace{1mm} = r_{k_1k_2} \Gamma_{k_1}\Gamma_{k_2}\Psi \vspace{2mm}\\
\Rightarrow \vspace{2mm} \\
\mathfrak{so}(2n) \hspace{2mm}\mapsto  \hspace{2mm}\mathfrak{su}(n).
\end{array}\end{equation}
\noindent Explicitly, a generic element of the surviving $\mathfrak{su}(n)$ subalgebra may be written in terms of raising and lowering operators as
\begin{equation}\begin{array}{lll}\label{sunladder}
\ell_n  &=&R_{j_1j_2}\left(a_{j_1}a^{\dagger}_{j_2} - a_{j_2}a^{\dagger}_{j_1}\right)\vspace{2mm}\\
&+&R_{j_1j_2}'i\left(a_{j_1}a^{\dagger}_{j_2} + a_{j_2}a^{\dagger}_{j_1}\right)\vspace{2mm}\\
&+& R_{j}\hspace{.5mm}i\left(a_{j}a^{\dagger}_{j}-a_{j+1}a^{\dagger}_{j+1}\right),
\end{array}\end{equation}
\noindent where $j_1\neq j_2$ and $R_{j_1j_2}, R_{j_1j_2}', R_j\in\mathbb{R}$.
\noindent In terms of $\{\Gamma_j\}$, these same $\mathfrak{su}(n)$ elements may be represented as 
\begin{equation}\begin{array}{lll}\label{sungamma}
\ell_n  &=& r_j \hspace{.5mm}\Gamma_{j}\Gamma_{j+n} \vspace{2mm}\\
&+& r_{j_1j_2}\left(\Gamma_{j_1}\Gamma_{j_2}+ \Gamma_{j_1+n}\Gamma_{j_2+n}\right)\vspace{2mm}\\
&+& r_{j_1j_2}'\left(\Gamma_{j_1}\Gamma_{j_2+n}+ \Gamma_{j_2}\Gamma_{j_1+n}\right),
\end{array}\end{equation}
\noindent where $r_j, r_{j_1j_2},r_{j_1j_2}'\in\mathbb{R},$ and $\sum_{j=1}^n r_j = 0.$ 

\subsection{Significance}

\noindent This result is significant for a number of reasons.  

(+)  It shows that a single consistency condition, (\ref{svconstraint}), can offer an answer to both of the open challenges 2. and 3. posed by Baez and Huerta in Section \ref{aguts}.  

(+)  As will appear later in this text, it eliminates the persistent  unwanted B-L symmetry from~\citep{fh2}.

(+)  We will also see that in the physically interesting cases of 
$$ \begin{array}{rcl} \textup{Spin(10)} &\mapsto&  \textup{SU(5),} \vspace{2mm}\\

 \textup{Pati-Salam} &\mapsto&  \textup{Standard Model,}  \vspace{2mm}\\

 \textup{Left-Right Symmetric} &\mapsto&  \textup{Standard Model,}
\end{array}$$
\noindent this consistency condition supplies a natural explanation for the Standard Model's maximal chirality.

\section{Simultaneous group actions of the reflective type}

\subsection{Motivation for considering $\RCHO$}

Recent work,~\citep{fh1}, \citep{fh2}, introduced the result that one generation of unconstrained fermions  can be identified with one copy of $\RCHO$,
$$\textup{one generation} \hspace{2mm}\leftrightarrow \hspace{2mm}\RCHO.$$
Furthermore, the division algebraic substructure of this algebra (surprisingly) led to a cascade of symmetry breakings in well-known particle models: 
$$\begin{array}{l}\textup{Spin(10)} \mapsto \textup{Pati-Salam} \mapsto \textup{Left-Right Symmetric}\mapsto\vspace{2mm}\\ 
\hspace{.75cm} \textup{Pre-Higgs Standard Model (+ B-L)}\mapsto \vspace{2mm}\\
\hspace{1.5cm} \textup{Post-Higgs Standard Model (+ B-L).}\end{array}$$

However, instead of identifying one generation with $\RCHO$, we will now construct one generation as a minimal left ideal within $\RCHO$'s \it multiplication algebra, \rm  $$\textup{one generation} \hspace{2mm}\leftrightarrow \hspace{2mm}\textup{MLI}.$$
\noindent Doing so will allow us to implement the multivector condition from the previous section.

$\RCHO$'s multiplication algebra, to be defined shortly, is isomorphic to its complex endomorphisms, $End\left(\RCHO\right)$.  It is also isomorphic as a matrix algebra to the complex Clifford algebra $\CLten$.  Hence it provides the facilities to build up a Spin(10) model, and due to its division algebraic substructure, the facilities to break that model down.

\subsection{The algebra $\RCHO$}

The algebra $\RCHO$ is also known as the Dixon algebra, due to its independent implementation in early particle models by Dixon.  Readers are encouraged to see~\citep{Dixon_recent} for an alternative perspective.  

Throughout this article, all tensor products will be assumed to be over $\mathbb{R}$ unless otherwise stated.  We write the standard $\mathbb{R}$-basis for $\mathbb{H}$ as $\{ \epsilon_0, \epsilon_1, \epsilon_2, \epsilon_3\}$, where $\epsilon_0 = 1$, $\epsilon_j^2 = -1$ for $j \in \{1, 2, 3\}$, and $\epsilon_1\epsilon_2 = \epsilon_3$, with cyclic permutations. 
These relations may be written more succinctly as $\epsilon_i\epsilon_j = -\delta_{ij}+\varepsilon_{ijk}\hspace{0.5mm}\epsilon_k$ for $i,j,k\in \{1,2,3\}$, where $ \varepsilon_{ijk}$ is the usual totally anti-symmetric tensor with $ \varepsilon_{123}=1$.  

Similarly, we write the standard $\mathbb{R}$-basis for $\mathbb{O}$ as $\{e_0, e_1, \dots, e_7\}$, where $e_0 = 1$ and $e_ie_j = -\delta_{ij}+f_{ijk}\hspace{0.5mm}e_k$  for $i,j,k\in \{1,2,\dots 7\}$.  Here, $ f_{ijk}$ is a totally anti-symmetric tensor with $f_{ijk}=1$ when $ijk\in \{124, 235, 346, 457, 561, 672, 713\}$.   The remaining values of $f_{ijk}$ are determined by anti-symmetry, and vanish otherwise.  We let $i$ denote a complex imaginary unit as usual. 

Let $\A := \RCHO = \CHO$. Note that $\A$ is naturally an algebra over $\mathbb{C}$: scalar multiplication by $c \in \C$ is defined by setting $c(x \otimes y \otimes w) = cx \otimes y \otimes w$ for $x \in \C, y \in \mathbb{H}, w \in \mathbb{O}$.  Multiplication of elements is defined by setting $(x_1 \otimes y_1 \otimes w_1)(x_2 \otimes y_2 \otimes w_2) = x_1x_2 \otimes y_1y_2 \otimes w_1w_2$ for all $x_1, x_2 \in \C, y_1, y_2 \in \mathbb{H}$, and $z_1, z_2 \in \mathbb{O}$. A $\C$-basis for $\A$ is given by $\{1 \otimes \epsilon_{\mu} \otimes e_{\nu} \mid \mu \in \{0, 1, 2, 3\}, \nu \in \{0, 1, \dots, 7\}\}$.  We see that $\A$ is a  32\hspace{0.5mm}$\C$-dimensional non-commutative, non-associative algebra. 

From now on, we identify $\C$, $\mathbb{H}$, and $\mathbb{O}$ with their images in $\A$ under the natural inclusion maps, thus writing $c$ instead of $c \otimes \epsilon_0 \otimes e_0$ for any $c \in \C$, writing $\epsilon_{\mu}$ for $1 \otimes \epsilon_{\mu} \otimes e_0$, and writing $e_{\nu}$ for $1 \otimes \epsilon_0 \otimes e_{\nu}$ (here $\mu \in \{0, 1, 2, 3\}, \hspace{1mm}\nu \in \{0, 1, \dots, 7\}$). Arbitrary elements of $\A$ are then written as $\sum_{\mu, \nu} c_{\mu\nu}\epsilon_{\mu}e_{\nu}$, where $c_{\mu\nu}\in\C,$ and we see that $\epsilon_{\mu}e_{\nu} = e_{\nu}\epsilon_{\mu}$ for all $\mu, \nu$. 

For those less comfortable with formal definitions, an example may help.  Suppose $a,b\in\A$ with  $a=4 \epsilon_1 e_2$ and $b=\left(5+i\right) \epsilon_2 e_4$.  Then $ab = \left(20+4i\right)\epsilon_3e_1.$

\subsection{From $\RCHO$ to its space of linear operators}

In works dating back as early as~\citep{GGquarks}, it was clear that the $\mathfrak{su}(3)$ adjoint part of gluons would be destined to reside not in $\mathbb{O},$ but rather in the space of linear operators on $\mathbb{O}$.  This space of linear operators can be realized as $\A$'s \it multiplication algebra.\rm

If bosons are to reside within $\A$'s multiplication algebra, then why not include fermions there as well?  One advantage of including fermions within $\A$'s multiplication algebra is that, unlike with $\A$, this space is now large enough to comfortably accommodate three generations.  Hence, we will position these fermions within $\A$'s multiplication algebra, in the form of minimal left ideals.

First let us define what is meant by \it multiplication algebra. \rm Suppose $x,y$ are elements in an algebra $\mathbb{D}.$  We then define $L_x(y):=xy$.  Similarly, define $R_x(y):=yx$.  Both $L_x$ and $R_x$ may then be seen as (possibly distinct) linear maps sending $y$ to some new element in $\mathbb{D}.$  Hence, $L_x$ and $R_x$ $\in End(\mathbb{D}).$

For $x,y,z\in\mathbb{D},$ these linear maps may be composed as in $L_x\circ L_y(z) = x(y(z))$, $R_x\circ R_y(z) = ((z)y)x,$ $L_x\circ R_y(z) = x((z)y),$ $R_x\circ L_y(z) = (y(z))x$, etc.  \emph{Composition} of these maps, $\circ$, defines an \emph{associative} multiplication rule.  Furthermore, maps may be added in the obvious way.  For example, $(L_x + L_y)(z)=xz+yz.$

We define the left multiplication algebra of $\mathbb{D}$ as the subalgebra of $End(\mathbb{D})$ generated by the $\{L_y\hspace{.5mm} | \hspace{.5mm}y\in\mathbb{D}\}$.  Similarly, the right multiplication algebra of $\mathbb{D}$ is the subalgebra of $End(\mathbb{D})$ generated by $\{R_y\hspace{.5mm} | \hspace{.5mm} y\in\mathbb{D}\}$.  We denote the left multiplication algebra of an algebra $\mathbb{D}$ as $\mathcal{L}_{\mathbb{D}},$ and its right multiplication algebra as $\mathcal{R}_{\mathbb{D}}.$  

As an example, consider the multiplication of two elements $A,B\in End(\A)$ with $A=4 L_{\epsilon_1e_2}$ and $B=\left(5+i\right) L_{\epsilon_2 e_4}$.  Then $AB:=A\circ B = \left(20+4i\right)L_{\epsilon_1e_2}\circ L_{\epsilon_2 e_4} = \left(20+4i\right)L_{\epsilon_3e_2}\circ L_{e_4}\neq  \left(20+4i\right)L_{\epsilon_3e_1}.$

\subsection{Clifford factors}

In this subsection, we examine the substructure of $\A$'s multiplication algebra.  We will find that it is this substructure that ultimately leads the Spin(10) model to splinter into some of its broken successors.

What are the multiplication algebras of $\C,$ $\mathbb{H},$ and $\mathbb{O}$?  Of $\RCHO$?  It can be confirmed that the following  isomorphisms hold (as matrix algebras): 
\begin{equation}\begin{array}{lllll} &\hspace{1mm}\hspace{1mm}&\mathcal{L}_{\C} = \mathcal{R}_{\C}&\hspace{2mm}\leftrightarrow\hspace{2mm}& Cl\left(0,1\right)\vspace{2mm}\\
 &\hspace{1mm}\hspace{1mm}&\mathcal{L}_{\mathbb{H}} &\hspace{2mm}\leftrightarrow\hspace{2mm}&Cl\left(0,2\right)\vspace{2mm}\\
 &\hspace{1mm}\hspace{1mm}&\mathcal{R}_{\mathbb{H}} &\hspace{2mm}\leftrightarrow\hspace{2mm}& Cl\left(0,2\right) \vspace{2mm}\\
 &\hspace{1mm}\hspace{1mm}&  \mathcal{L}_{\mathbb{O}}\simeq \mathcal{R}_{\mathbb{O}} &\hspace{2mm}\leftrightarrow\hspace{2mm}& Cl\left(0,6\right).\\
\end{array}\end{equation}
Since $\C$ is abelian,  $\mathcal{L}_{\C}$ and $\mathcal{R}_{\C}$ are equal elementwise.  That is, $L_{c} = R_{c}\hspace{2mm}\forall c\in \C$.  We identify these linear maps with $Cl(0,1),$ where the vector generating $Cl(0,1)$ is given by multiplication by the complex imaginary unit, $L_i=R_i$. 

Unlike the complex numbers, $\mathbb{H}$ is non-abelian, and perhaps unsurprisingly, $\mathcal{L}_{\mathbb{H}}$ and $\mathcal{R}_{\mathbb{H}}$ typically provide distinct linear maps on $\mathbb{H}$.  We may identify each of $\mathcal{L}_{\mathbb{H}}$ and $\mathcal{R}_{\mathbb{H}}$ with one copy of $Cl(0,2)$, thereby identifying the combined multiplication algebras of $\mathbb{H}$ with $Cl\left(0,2\right)\otimes Cl\left(0,2\right).$  For concreteness, we will take $L_{\epsilon_1}$ and $L_{\epsilon_2}$ to generate the first copy of $Cl(0,2),$ and $R_{\epsilon_1}$ and $R_{\epsilon_2}$ to generate the second copy of $Cl(0,2),$ although there is clearly a continuum of equivalent choices.  

Finally, the left- and right-multiplication algebras of $\mathbb{O}$ may each be identified with the Clifford algebra $Cl(0,6)$.  However, unlike with the quaternions, it is possible to show that each element of $\mathcal{R}_{\mathbb{O}}$ gives the same linear map on $\mathbb{O}$ as some element in $\mathcal{L}_{\mathbb{O}}.$  For example,  $\forall f\in\mathbb{O}$, 
\begin{equation}\begin{array}{lll}
R_{e_7}f:= fe_7 &=& \frac{1}{2} \left(  -e_7f + e_1\left(e_3 f\right)+ e_2\left(e_6 f\right)+ e_4\left(e_5 f\right)   \right)\vspace{2mm}\\
&=& \frac{1}{2} \left(  -L_{e_7} + L_{e_1}L_{e_3} + L_{e_2}L_{e_6} + L_{e_4}L_{e_5} \right)f,\vspace{2mm}
\end{array}\end{equation}
\noindent where $e_1, e_2, \dots e_7$ represent octonionic imaginary units. Therefore we see that although $L_{a}$ and $R_{a}$ are not equal for every $a\in\mathbb{O}$, $\mathcal{L}_{\mathbb{O}}$ and $\mathcal{R}_{\mathbb{O}}$ do provide the same set of linear maps on $\mathbb{O}.$  Hence, we write that $\mathcal{L}_{\mathbb{O}}\simeq \mathcal{R}_{\mathbb{O}}.$  The set of octonionic linear maps, identified with $Cl(0,6),$ may be generated by $\{L_{e_j}\}$, or equivalently by $\{R_{e_j}\},$ where $j=1,\dots ,6.$  Again, we emphasize that a continuum of equivalent choices exists (6-sphere $S^6$) .

In short, we find that the linear maps coming from the multiplication algebras of $\C,$ $\mathbb{H},$ and $\mathbb{O}$ are given by $Cl(0, 1),$ $Cl(0,2)\otimes Cl(0,2),$ and $Cl(0,6 )$ respectively.  The gradings for these Clifford algebras can result from choosing the Clifford algebra's generators to be multiplication by an orthogonal set of imaginary units.  $Cl(0,1)$ may be generated by $L_i=R_i;$ the two factors of $Cl(0,2)\otimes Cl(0,2)$ may be generated by $\{L_{\epsilon_{1}}, L_{\epsilon_{2}}\}$ and $\{R_{\epsilon_{1}}, R_{\epsilon_{2}}\};$  $Cl(0,6)$ may be generated by $\{L_{e_{j}}\}$, or alternatively by $\{R_{e_{j}}\},$ where $j\in\{1,\dots ,6\}.$

Finally, combining $\mathcal{L}_{\C} = \mathcal{R}_{\C},$ $\mathcal{L}_{\mathbb{H}},$  and $\mathcal{L}_{\mathbb{O}} \simeq \mathcal{R}_{\mathbb{O}},$ gives $\RCHO$'s left multiplication algebra as
\begin{equation} \begin{array}{l}\label{factor}
\mathcal{L}_{\C}\otimes \mathcal{L}_{\mathbb{H}} \otimes \mathcal{L}_{\mathbb{O}} \vspace{2mm}\\
\simeq Cl(0,1)\otimes Cl(0,2)  \otimes Cl(0,6) \simeq \CLeight .
\end{array}\end{equation}

Combining $\mathcal{L}_{\C} = \mathcal{R}_{\C},$ $\mathcal{L}_{\mathbb{H}},$ $\mathcal{R}_{\mathbb{H}},$ and $\mathcal{L}_{\mathbb{O}} \simeq \mathcal{R}_{\mathbb{O}},$ gives $\RCHO$'s full multiplication algebra as
\begin{equation} \begin{array}{l}\label{factor}
\mathcal{L}_{\C}\otimes \mathcal{L}_{\mathbb{H}}\otimes \mathcal{R}_{\mathbb{H}} \otimes \mathcal{L}_{\mathbb{O}} \vspace{2mm}\\
\simeq Cl(0,1)\otimes Cl(0,2) \otimes Cl(0,2) \otimes Cl(0,6) \simeq \CLten .\vspace{2mm}
\end{array}\end{equation}

\subsection{Hermitian conjugation}

We define hermitian conjugation, $\dagger$, on elements of $End(\A)$ as the involution such that $L_{e_j}^{\dagger} = -L_{e_j}$ $\forall j\in\{1, 2,\dots 7\}$, $L_{\epsilon_m}^{\dagger} = -L_{\epsilon_m}$ and $R_{\epsilon_m}^{\dagger} = -R_{\epsilon_m}$ $\forall m\in\{1, 2,3\}$, and $L_i^{\dagger} = -L_i.$  As with matrix algebras, the hermitian conjugate defined here obeys $\left(ab\right)^{\dagger} = b^{\dagger} a^{\dagger}$ $\forall a,b\in End\left(\A\right).$

\subsection{Division algebraic reflections\label{dar}}

We have now explored the substructure of $\RCHO$'s multiplication algebra.  Next, we would  like to know how one goes about putting this substructure to use.  

In short, we will employ the $\mathbb{Z}_2$-gradings that are inherent to each Clifford algebraic factor in equation~(\ref{factor}).  Or, said another way, we will make use of  generalized notions of \it reflection. \rm

Referring to the first factor in equation (\ref{factor}), we know that $\mathcal{L}_{\C}$ is isomorphic to the Clifford algebra $Cl(0,1)$.  In this case, complex conjugation, denoted $*_{\overrightarrow{\mathbb{C}}}\vspace{.5mm}$, maps $L_i\mapsto -L_i,$ and defines the $\mathbb{Z}_2$-grading on this Clifford algebra.  It also induces a reflection of any complex number across the real axis.

It is straightforward to identify analog reflections in the subsequent  Clifford algebraic factors of equation~(\ref{factor}).  For example, a $\mathbb{Z}_2$-grading on $\mathcal{L}_{\mathbb{H}}\simeq Cl(0,2)$ may be induced by a quaternionic reflection $*_{\overrightarrow{\mathbb{H}}}:L_{\epsilon_m} \mapsto -L_{\epsilon_m}$  for $m\in\{1,2\}$.  Similarly, a $\mathbb{Z}_2$-grading on $\mathcal{L}_{\mathbb{O}}\simeq Cl(0,6)$ may be induced by an octonionic reflection $*_{\overrightarrow{\mathbb{O}}}: L_{e_j} \mapsto -L_{e_j}$ for $j\in\{1,2, \dots 6\}$.

Certain attentive readers might rightfully question why these octonionic and quaternionic reflections involve only six and two imaginary units respectively.  Why not seven and three?  This comes from the left multiplication algebras being  isomorphic to $Cl(0,6)$ and $Cl(0,2)$, not $Cl(0,7)$ and $Cl(0,3)$.  Nonetheless, rest assured that the final imaginary units become reintegrated when these reflections are composed with Clifford algebraic reversion:  $(\Gamma_i \Gamma_j)^{\rho} = \Gamma_j \Gamma_i.$  It is tedious, but straightforward, to confirm that $L_{e_7} = L_{e_1}L_{e_2}L_{e_3}L_{e_4}L_{e_5}L_{e_6}$.  Hence, $(L_{e_7}^{*_{\overrightarrow{\mathbb{O}}}})^{{\rho}} = L_{e_7}^{{\rho}} = -L_{e_7}.$  Similarly, $L_{\epsilon_3} = L_{\epsilon_1}L_{\epsilon_2}$.  Hence, $(L_{\epsilon_3}^{*_{\overrightarrow{\mathbb{H}}}})^{{\rho}} = L_{\epsilon_3}^{{\rho}} = -L_{\epsilon_3}.$

For ease of exposition, we have chosen specific imaginary units above.  However, readers should be aware that an entire 6-sphere of equivalent choices exist for $\mathcal{L}_{\mathbb{O}}$, a 2-sphere of equivalent choices exist for $\mathcal{L}_{\mathbb{H}}$, and a  0-sphere ($\mathbb{Z}_2$) of equivalent choices exists for $\mathcal{L}_{\C}$.

\subsection{An explicit division algebraic representation of $\CLten$}

There are many ways to generate $\CLten \simeq End(\A)$ so that the results in this paper materialize.  One such set is given by
\begin{equation}\label{DivAlgGammas}
\Gamma_j = iL_{e_j}L_{e_7}R_{\epsilon_1}, \hspace{.5cm} \Gamma_{m+6} = L_{e_7}L_{\epsilon_m}R_{\epsilon_1}, \hspace{.5cm} \Gamma_{10}=R_{\epsilon_2}
\end{equation}
\noindent for $j\in\{1,\dots 6\},$ and $m\in\{1,2, 3\}.$  (Another, possibly simpler, starting point has been found recently by John Barrett).  We define 
\begin{equation}\Gamma_{11}:= \prod_{p=1}^{10}\Gamma_p, 
\end{equation}
\noindent which in this case becomes
\begin{equation}\Gamma_{11}=-R_{\epsilon_3}.
\end{equation}
\noindent It allows us to define a chirality operator $-i\Gamma_{11}=iR_{\epsilon_3}$.  We specify raising operators $a_c$ and lowering operators $a_c^{\dagger}$ as
\begin{equation}  \begin{array}{lll} \label{ladders}
a_1^{\dagger}:= \frac{1}{2}(\Gamma_5 +i\Gamma_4) & \hspace{5mm} &a_1:= \frac{1}{2}(-\Gamma_5 +i\Gamma_4) \vspace{2mm}\\

a_2^{\dagger}:= \frac{1}{2}(\Gamma_3 +i\Gamma_1) & \hspace{5mm} &a_2:= \frac{1}{2}(-\Gamma_3 +i\Gamma_1) \vspace{2mm}\\

a_3^{\dagger}:= \frac{1}{2}(\Gamma_6 +i\Gamma_2) & \hspace{5mm} &a_3:= \frac{1}{2}(-\Gamma_6 +i\Gamma_2) \vspace{2mm}\\

a_4^{\dagger}:= \frac{1}{2}(\Gamma_8 +i\Gamma_7) & \hspace{5mm} &a_4:= \frac{1}{2}(-\Gamma_8 +i\Gamma_7) \vspace{2mm}\\

a_5^{\dagger}:= \frac{1}{2}(\Gamma_{10} +i\Gamma_9) & \hspace{5mm} &a_5:= \frac{1}{2}(-\Gamma_{10} +i\Gamma_9), \vspace{2mm}\\
\end{array}\end{equation}
\noindent where readers may notice that we have made a permissible relabeling of indices relative to equation~(\ref{a2g}) in order to connect with previous work,~\citep{fh1}, \citep{fh2}.  Defining a nilpotent object
\begin{equation}  \Omega := \prod_{c=1}^5 a_c
\end{equation}
\noindent then sets our hermitian vacuum state as  $v:=\Omega^{\dagger}\Omega$.  Finally, we construct a minimal left ideal as 
\begin{equation} \Psi := \CLten v.
\end{equation}
\noindent This gives a $32\hspace{.5mm} \C$ dimensional subspace of $\CLten$.  For this first article in the series, we will be particularly interested in the $16\hspace{.5mm} \C$ semi-spinor given by 
\begin{equation}
\Psi_{\textup{L}}:=\frac{1}{2}(1+i\Gamma_{11})\Psi.
\end{equation}
\noindent In the second article of this series,~\citep{fr2}, we will then extend the semi-spinor in two inequivalent ways.

\subsection{Cascade of particle symmetries:  \\the $\mathfrak{spin}$(10) line}

\begin{figure}[h!]
\begin{center}
\includegraphics[width=9cm]{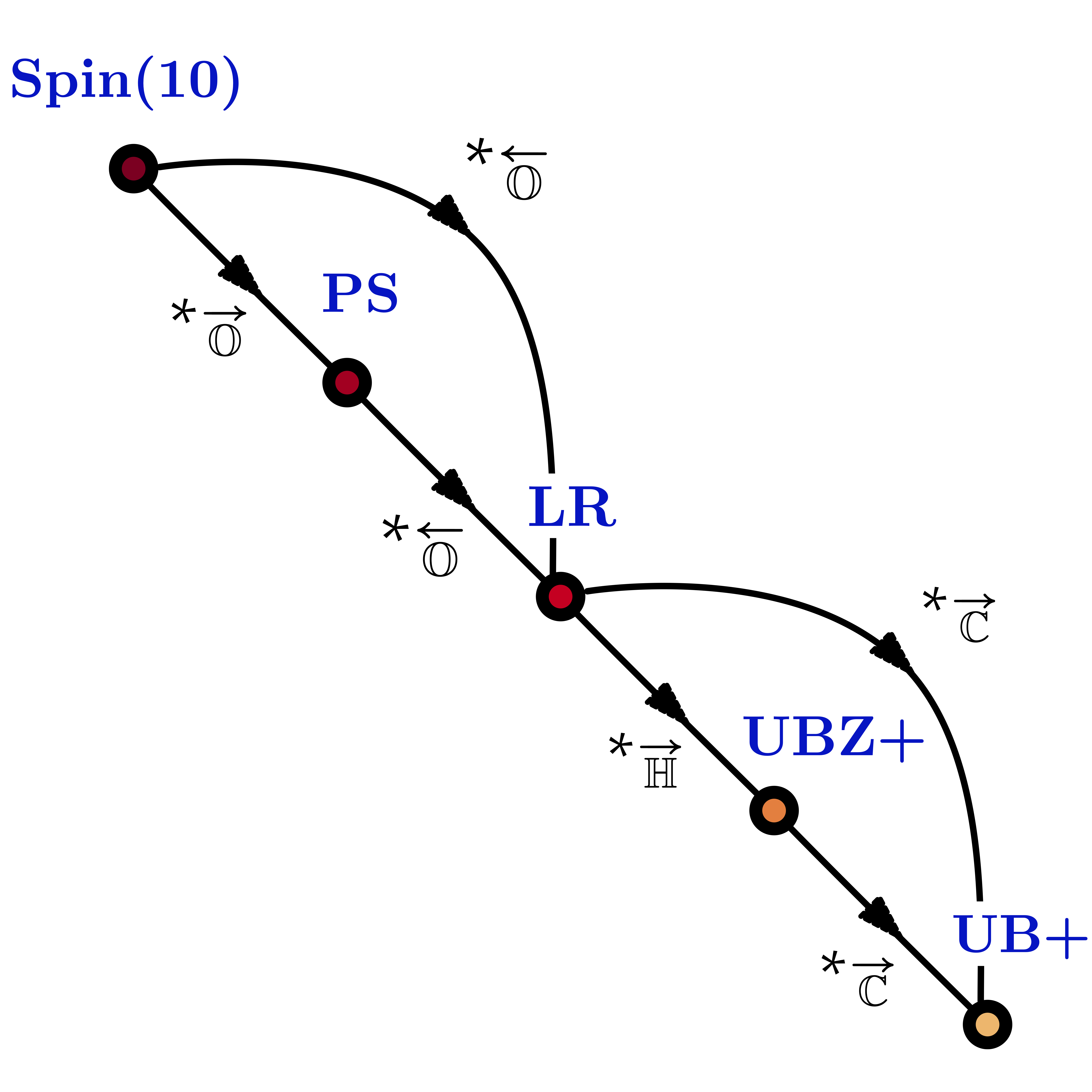}
\caption{\label{casc1}
Four types of division algebraic reflections are relevant in the left-multiplication space.  Invariance under the first octonionic reflection restricts Spin(10) to Pati-Salam (PS).  Invariance under the second octonionic reflection restricts Pati-Salam to Left-Right Symmetric (LR).  Invariance under a quaternionic reflection restricts Left-Right Symmetric to the Standard Model's unbroken  symmetries, together with a $\mathfrak{u}(1)$ related to B-L, and another $\mathfrak{u}(1)$ related to the $Z^0$ boson, ({\textup{UBZ+}}).  Invariance under complex conjugation eliminates the $\mathfrak{u}(1)$ related to the $Z^0$, leaving the Standard Model's unbroken  symmetries, together with B-L, (UB+).    Readers can confirm that certain paths bifurcate.  }
\end{center}
\end{figure}

With our fermions now embedded within $End(\A)$ we will demonstrate how division algebraic reflections prompt them to fragment.  Please see Figure~(\ref{casc1}).

A generic element of the 45-dimensional $\mathfrak{spin}$(10) may be written as real linear combinations of bivectors:
\begin{equation}\begin{array}{ll}\label{so10D}
\ell_{10}^D:=&r_{ij}L_{e_i}L_{e_j} + r_m L_{\epsilon_m} + r'_m L_{\epsilon_m}L_{e_7}R_{\epsilon_3} \vspace{2mm}
\\&+\hspace{1mm} r_{mj}' iL_{\epsilon_m}L_{e_j} + r_j''iL_{e_j}L_{e_7}R_{\epsilon_3}
\end{array}\end{equation}

\noindent for $i,j\in\{1,\dots 6\}$ with $i\neq j,$  $m\in\{1,2, 3\},$ and $r_{ij}, r_m, r'_m, r_{mj}', r_j''\in\R$.  Here the `D' in $\ell_{10}^D$ refers to `Dirac'.

When applied to our semi-spinor $\Psi_{\textup{L}}$, this $\mathfrak{spin}$(10) action simplifies considerably:
\begin{equation}\begin{array}{lc}
\ell_{10}^D\Psi_{\textup{L}} =& (r_{ab}L_{e_a}L_{e_b} + r'_{mn} L_{\epsilon_m}L_{\epsilon_n} + r_{ma}'' iL_{\epsilon_m}L_{e_a} )\Psi_{\textup{L}},\vspace{2mm}\\
&\mathfrak{spin}(10) \hspace{.7cm}
\end{array}\end{equation}
\noindent where this time $a,b\in\{1,\dots7\}$ with $a\neq b,$ and $m,n\in\{1,2,3\}$ with $m\neq n,$ while $r_{ab},  r'_{mn}, r_{ma}''\in\R.$  This Weyl action  matches that found earlier in~\citep{fh2}.  Let us then define
\begin{equation}\begin{array}{lc}\label{W10}
\ell_{10}:=& r_{ab}L_{e_a}L_{e_b} + r'_{mn} L_{\epsilon_m}L_{\epsilon_n} + r_{ma}'' iL_{\epsilon_m}L_{e_a} .\vspace{2mm}\\
&\mathfrak{spin}(10) \hspace{.7cm}
\end{array}\end{equation}

With this representation of $\mathfrak{spin}(10)$ at our disposal, we may now make use of the division algebraic reflections introduced in Section~\ref{dar}.  That these reflections would reduce the SO(10) grand unified theory to other well-known particle models came originally as a surprise.

It is important to note that the $\mathfrak{spin}(10)$ element $\ell_{10}$ lives exclusively in the \it left\rm-multiplication algebra of $\A$.  Therefore there are four types of division algebraic reflections that will be relevant for us:  those related to $\langle 1\rangle$ $\mathcal{L}_{\mathbb{O}},$  $\langle 2\rangle$ $\mathcal{R}_{\mathbb{O}},$  $\langle 3\rangle$ $\mathcal{L}_{\mathbb{H}},$ and $\langle 4\rangle$ $\mathcal{L}_{\mathbb{C}}.$

$\langle 1\rangle$  First of all, we would like to know:  Which Lie subalgebra of our $\mathfrak{spin}(10)$ symmetries is invariant under an octonionic reflection?  Readers can easily confirm that setting 
\begin{equation}
{\ell}_{10}\Psi_{\textup{L}} = {\ell}_{10}^{*_{\overrightarrow{\mathbb{O}}}}\Psi_{\textup{L}} \hspace{1cm} \forall\Psi_L
\end{equation}
\noindent restricts $\mathfrak{spin}(10)$'s $\ell_{10}$ to the Pati-Salam symmetries, $\ell_{\textup{PS}}:$
\begin{equation}\begin{array}{rll}
{\ell}_{\textup{PS}}\Psi_{\textup{L}} = (r_{ij}L_{e_i}L_{e_j} &+& r_m L_{\epsilon_m} + r'_m iL_{\epsilon_m}L_{e_7})\Psi_{\textup{L}} \vspace{2mm}\\
\mathfrak{spin}(6) &\oplus& \mathfrak{spin}(4)
\end{array}\end{equation}
\noindent for $i,j\in\{1,\dots 6\}$ with $i\neq j,$ and $m\in\{1,2, 3\}.$  Note that $\mathfrak{spin}(6) = \mathfrak{su}(4),$ and $\mathfrak{spin}(4) = \mathfrak{su}(2)\oplus\mathfrak{su}(2)$.

Next, we will consider those symmetries furthermore invariant under an analogous octonionic reflection, deriving from \it right \rm multiplication.  Namely $*_{\overleftarrow{\mathbb{O}}}: R_{e_j}\mapsto -R_{e_j}$ $\forall j \in \{1, \dots, 6\}.$   

$\langle 2\rangle$  Which Lie subalgebra of the Pati-Salam symmetries is invariant under this complementary octonionic reflection? With a little work, readers will find the Lie algebra of the Left-Right Symmetric model.  Setting 
\begin{equation}
{\ell}_{\textup{PS}}\Psi_{\textup{L}} = {\ell}_{\textup{PS}}^{*_{\overleftarrow{\mathbb{O}}}}\Psi_{\textup{L}}\hspace{1cm} \forall \Psi_L 
\end{equation}
\noindent restricts $\ell_{\textup{PS}}$ to the Left-Right Symmetric model's gauge symmetries, $\ell_{\textup{LR}}:$ 
\begin{equation}\begin{array}{rll}
{\ell}_{\textup{LR}}\Psi_{\textup{L}} = (r_{ij}''L_{e_i}L_{e_j} &+& r_m L_{\epsilon_m} + r'_m iL_{\epsilon_m}L_{e_7})\Psi_{\textup{L}}, \vspace{2mm}\\
\mathfrak{u}(3) &\oplus&  \mathfrak{spin}(4)
\end{array}\end{equation}
\noindent where the coefficients $r_{ij}''\in\R$ are this time restricted to give a $\mathfrak{u}(3)$ subalgebra of $\mathfrak{spin}(6)$.  Explicitly, we have eight $\mathfrak{su}(3)_C$ generators 
\begin{equation}\begin{array}{lll}\label{su3}
i\Lambda_1:=\frac{1}{2}\left(L_{34}-L_{15} \right) &\hspace{1mm} &
i\Lambda_2:= \frac{1}{2}\left(L_{14}+L_{35} \right)  \vspace{2mm}  \\

i\Lambda_3:=\frac{1}{2}\left(L_{13}-L_{45} \right)&\hspace{1mm} &
i\Lambda_4:= -\frac{1}{2}\left(L_{25}+L_{46} \right)  \vspace{2mm}  \\

i\Lambda_5:=\frac{1}{2}\left(L_{24}-L_{56} \right) &\hspace{1mm} &
i \Lambda_6:=-\frac{1}{2}\left(L_{16}+L_{23} \right)\vspace{2mm}
  \\  
  
i\Lambda_7:= - \frac{1}{2}\left(L_{12}+L_{36} \right)&\hspace{1mm} &  
i \Lambda_8 :=\frac{-1}{2\sqrt{3}}\left(L_{13}+L_{45}  -  2L_{26} \right),  
\end{array}\end{equation}
\noindent where $L_{ij}$ is shorthand for the octonionic $L_{e_i}L_{e_j}$.  The generator for the $\mathfrak{u}(1)_{B-L}$ subalgebra of $\mathfrak{u}(3)$ may be described as
\begin{equation}i(B-L) := \frac{1}{3}(L_{13}+L_{26}+L_{45}).
\end{equation}
\noindent An alternative route was found by Boyle in~\citep{boyle1} that maps Spin(10) directly to the Left-Right Symmetric model via $*_{\overleftarrow{\mathbb{O}}}$ in the context of the exceptional Jordan algebra.  This result likewise holds in our current model.

$\langle 3\rangle$ Next, fixing 
\begin{equation}
{\ell}_{\textup{LR}}\Psi_{\textup{L}} = {\ell}_{\textup{LR}}^{*_{\overrightarrow{\mathbb{H}}}}\Psi_{\textup{L}}\hspace{1cm} \forall \Psi_L
\end{equation}
\noindent restricts ${\ell}_{\textup{LR}}$ to a $\mathfrak{u}(3)\oplus\mathfrak{u}(1)\oplus\mathfrak{u}(1)$ subalgebra, which we will describe by elements ${\ell}_{{\textup{UBZ+}}}$:
\begin{equation}\begin{array}{rllrll}\label{ubz+}
{\ell}_{{\textup{UBZ+}}}\Psi_{\textup{L}} = (r_{ij}''L_{e_i}L_{e_j} &+& r_3 L_{\epsilon_3} &+& r'_3 iL_{\epsilon_3}L_{e_7})\Psi_{\textup{L}}.  \vspace{2mm}\\
\mathfrak{u}(3) &\oplus&  \mathfrak{u}(1)&\oplus&  \mathfrak{u}(1)
\end{array}\end{equation}
\noindent This less recognizable Lie subalgebra describes the Standard Model's unbroken gauge symmetries, together with a B-L symmetry, and an additional $\mathfrak{u}(1)$ symmetry corresponding to the $Z^0$ boson.

$\langle 4\rangle$ Finally, requiring this action's invariance under complex conjugation, 
\begin{equation}
{\ell}_{{\textup{UBZ+}}}\Psi_{\textup{L}} = {\ell}_{{\textup{UBZ+}}}^{*_{\overrightarrow{\mathbb{C}}}}\Psi_{\textup{L}} \hspace{1cm} \forall \Psi_L
\end{equation}
\noindent restricts ${\ell}_{{\textup{UBZ+}}}$ to a $\mathfrak{u}(3)\oplus\mathfrak{u}(1)$ Lie subalgebra, described by elements ${\ell}_{{\textup{UB+}}}$:

\begin{equation}\begin{array}{rll}\label{ub+}
{\ell}_{{\textup{UB+}}}\Psi_{\textup{L}} = (r_{ij}''L_{e_i}L_{e_j} &+& r_3 L_{\epsilon_3})\Psi_{\textup{L}},  \vspace{2mm}\\
\mathfrak{u}(3) &\oplus&  \mathfrak{u}(1)
\end{array}\end{equation}
\noindent thereby eliminating the symmetry corresponding to the $Z^0$ boson.  This leaves us with the Standard Model's unbroken gauge symmetries, after the Higgs mechanism, in addition to a B-L symmetry.

\subsection{Cascade of particle symmetries:  the Georgi-Glashow line}

\begin{figure}[h!]
\begin{center}
\includegraphics[width=8.2cm]{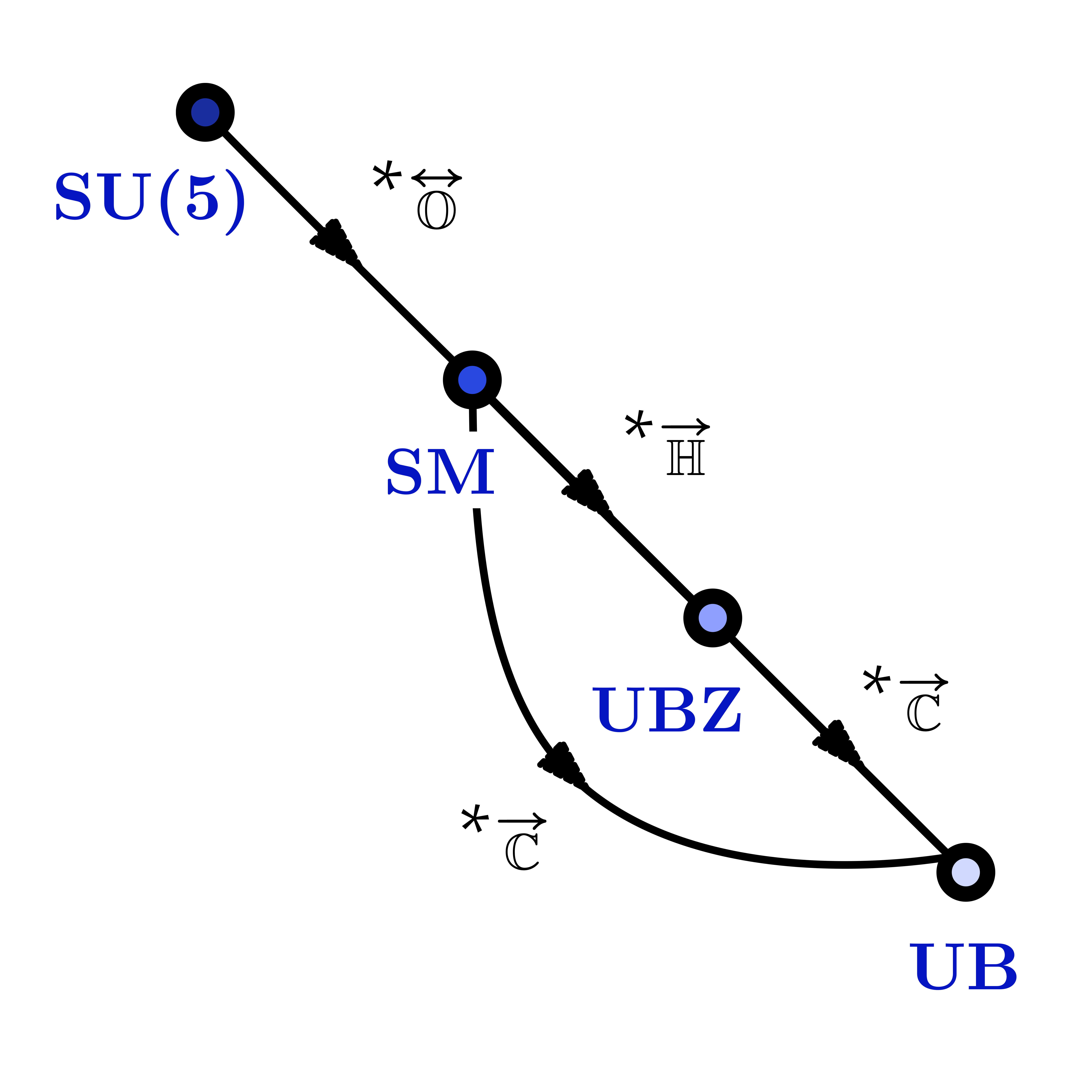}
\caption{\label{casc2}
The same division algebraic reflections as Figure (\ref{casc1}) may be applied to the Georgi-Glashow model.  We find that invariance under either of the octonionic reflections sends SU(5) to the Standard Model (SM).  Unlike with~\citep{fh2}, though, chirality now arises organically, and the unwanted B-L symmetry is eliminated.  From here, invariance under a quaternionic reflection sends the Standard Model's symmetries to the Standard Model's unbroken symmetries, together with a $\mathfrak{u}(1)$ related to the $Z^0$ boson, (UBZ).  Invariance under the complex reflection finally eliminates this $\mathfrak{u}(1)$ factor to give the Standard Model's familiar unbroken symmetries, (UB).  Again, complex conjugation provides a bypass that maps  SM directly to UB.}
\end{center}
\end{figure}

We will now mark a path of broken symmetries, parallel to that of the previous section.  This time, however, we will begin with Georgi and Glashow's SU(5) model.  Please see Figure~(\ref{casc2}).

The 24$\hspace{.5mm \R}$ dimensional  $\mathfrak{su}$(5) Lie algebra may be represented as
\begin{equation}\begin{array}{lll}\label{sunladder5}
\ell_{5}  &=&R_{cd}\left(a_{c}a^{\dagger}_{d} - a_{d}a^{\dagger}_{c}\right)\vspace{2mm}\\
&+&R_{cd}'\hspace{.5mm}i\left(a_{c}a^{\dagger}_{d} + a_{d}a^{\dagger}_{c}\right)\vspace{2mm}\\
&+& R_c\hspace{.5mm} i\left(a_{c}a^{\dagger}_{c}-a_{c+1}a^{\dagger}_{c+1}\right),
\end{array}\end{equation}
\noindent for $c,d \in\{1,\dots 5\}$ and $c\neq d$.  Here,  $R_{cd}, R_{cd}', R_c\in\mathbb{R}$.   This representation of $\mathfrak{su}(5)$ corresponds to the Lie subalgebra of $\mathfrak{spin}(10)$  surviving the multivector condition, as in equation~(\ref{sunladder}).

Substituting in the division algebraic raising and lowering operators of equation~(\ref{ladders}) into equation~(\ref{sunladder5}) gives an $\mathfrak{su}(5)$ element, $\ell_5$, acting on $\Psi_{\textup{L}}$ as
\begin{equation}\begin{array}{lllll}
 &&\ell_5 \Psi_{\textup{L}}=&& \vspace{2mm}\\
&&\big(r_j i \Lambda_j + r'_m L_{\epsilon_m}s  + irY && \vspace{2mm}\\
&&\hspace{.5mm}+\hspace{.7mm}it_1\left( L_{\epsilon_1}L_{e_4}+L_{\epsilon_2}L_{e_5}\right)&+&it_1'\left( L_{\epsilon_1}L_{e_5}-L_{\epsilon_2}L_{e_4}\right)\vspace{2mm}\\
&& \hspace{.5mm}+\hspace{.7mm}it_1''\left( L_{\epsilon_3}L_{e_4}-iL_{e_5}L_{e_7}\right)&+&it_1'''( L_{\epsilon_3}L_{e_5}+iL_{e_4}L_{e_7})\vspace{2mm}\\
&& \hspace{.5mm}+\hspace{.7mm}it_2\left( L_{\epsilon_1}L_{e_1}+L_{\epsilon_2}L_{e_3}\right)&+&it_2'( L_{\epsilon_1}L_{e_3}-L_{\epsilon_2}L_{e_1})\vspace{2mm}\\
&& \hspace{.5mm}+\hspace{.7mm}it_2''\left( L_{\epsilon_3}L_{e_1}-iL_{e_3}L_{e_7}\right)&+&it_2'''( L_{\epsilon_3}L_{e_3}+iL_{e_1}L_{e_7})\vspace{2mm}\\
&& \hspace{.5mm}+\hspace{.7mm}it_3\left( L_{\epsilon_1}L_{e_2}+L_{\epsilon_2}L_{e_6}\right)&+&it_3'( L_{\epsilon_1}L_{e_6}-L_{\epsilon_2}L_{e_2})\vspace{2mm}\\
&& \hspace{.5mm}+\hspace{.7mm}it_3''\left( L_{\epsilon_3}L_{e_2}-iL_{e_6}L_{e_7}\right)&+&it_3'''( L_{\epsilon_3}L_{e_6}+iL_{e_2}L_{e_7}) \big) \Psi_{\textup{L}}\vspace{2mm}\\
&&\hspace{1.7cm}\mathfrak{su}(5)&&
\end{array}\end{equation}

\noindent where $j\in\{1,\dots 8\},$ $m\in\{1,2,3\}$,  and $r_j, r'_m, r, t_k, t_k',\vspace{.5mm}$ $t_k'', t_k'''\vspace{.5mm}  \in \R$.  The idempotent $s$ is defined as $s:=\frac{1}{2}(1+iL_{e_7}),$ and preserves $\mathfrak{su}(2)_{\textup{L}}$-active states. The weak hypercharge generator, $iY$, is defined via 
\begin{equation}iY \Psi_{\textup{L}}:=\left( \frac{1}{6}\left(L_{13} +L_{26}+L_{45} \right) - \frac{1}{2}L_{\epsilon_3}s^*\right)\Psi_{\textup{L}},
\end{equation}
\noindent following the conventions of~\citep{BM}.

With our $\mathfrak{su}(5)$ action now established, division algebraic reflections may  be employed so as to break these symmetries sequentially.

$\langle 1 \rangle$ Which Lie subalgebra of Georgi and Glashow's $\mathfrak{su}(5)$ exhibits immunity to an octonionic reflection?  Readers may easily confirm that  fixing the constraint 
\begin{equation}{\ell}_{5}\Psi_{\textup{L}} = {\ell}_{5}^{*_{\overrightarrow{\mathbb{O}}}}\Psi_{\textup{L}} \hspace{1cm} \forall \Psi_L
\end{equation} 
\noindent restricts $\mathfrak{su}(5)$ to the Standard Model's symmetries, $\mathfrak{su}(3)\oplus\mathfrak{su}(2)\oplus\mathfrak{u}(1)$.  

$\langle 2 \rangle$ Interestingly, we find that the right-multiplication condition 
\begin{equation}{\ell}_{5}\Psi_{\textup{L}} = {\ell}_{5}^{*_{\overleftarrow{\mathbb{O}}}}\Psi_{\textup{L}} \hspace{1cm} \forall \Psi_L
\end{equation} 
has an identical effect.  In either case, we are left with the Standard Model's gauge symmetries acting on $\Psi_{\textup{L}}$ as 
\begin{equation}\begin{array}{rccll}\label{lsm}
\ell_{\textup{SM}}\Psi_{\textup{L}} = \big( r_j i \Lambda_j &+& r'_m L_{\epsilon_m}s  &+& irY\big)\Psi_{\textup{L}}.\vspace{2mm}\\
\mathfrak{su}(3)&\oplus&\mathfrak{su}(2)&\oplus&\mathfrak{u}(1)
\end{array}\end{equation}

It is worth taking a moment to point out that the result established here is not only obtaining the correct Standard Model Lie algebras.  Over the years, a recurring challenge in algebraic models has been to also secure the correct  chiral fermion \it representations. \rm  That the correct representations are indeed realized  here will be established in~\citep{fr2}.

$\langle 3 \rangle$  It is perhaps this next quaternionic step that is of most phenomenological interest.  The condition 
\begin{equation}{\ell}_{\textup{SM}}\Psi_{\textup{L}} = {\ell}_{\textup{SM}}^{*_{\overrightarrow{\mathbb{H}}}}\Psi_{\textup{L}} \hspace{1cm} \forall \Psi_L
\end{equation} 
\noindent can be seen to restrict $\mathfrak{su}(3)\oplus\mathfrak{su}(2)\oplus\mathfrak{u}(1)$ to $\mathfrak{su}(3)\oplus\mathfrak{u}(1)\oplus\mathfrak{u}(1)$.  This resulting $\mathfrak{su}(3)\oplus\mathfrak{u}(1)\oplus\mathfrak{u}(1)$ action describes the Standard Model's unbroken gauge symmetries, together with an additional $\mathfrak{u}(1)$ symmetry associated with the $Z^0$ boson.  Explicitly, its action on $\Psi_{\textup{L}}$ reduces to
\begin{equation}\begin{array}{rccll}\label{ubz}
\ell_{\textup{UBZ}}\Psi_{\textup{L}} = \big( r_j i \Lambda_j &+& r'_3 L_{\epsilon_3}s  &+& irY\big)\Psi_{\textup{L}}.\vspace{2mm}\\
\mathfrak{su}(3)&\oplus&\mathfrak{u}(1)&\oplus&\mathfrak{u}(1)
\end{array}\end{equation}
\noindent Its meaning for electroweak symmetry breaking will be discussed briefly later in this text.

$\langle 4 \rangle$ Finally we arrive at invariance under complex conjugation. Applying the constraint that 
\begin{equation}{\ell}_{\textup{UBZ}}\Psi_{\textup{L}} = {\ell}_{\textup{UBZ}}^{*_{\overrightarrow{\mathbb{C}}}}\Psi_{\textup{L}} \hspace{1cm} \forall \Psi_L
\end{equation} 
\noindent leaves us with the Standard Model's unbroken gauge symmetries, $\mathfrak{su}(3)\oplus\mathfrak{u}(1)$.  A generic element of this Lie algebra  acts on $\Psi_{\textup{L}}$ as
\begin{equation}\begin{array}{lrcl}
\ell_{\textup{UB}}\Psi_{\textup{L}} = &\big( r_j i \Lambda_j &+& r'iQ\big)\Psi_{\textup{L}},\vspace{2mm}\\
&\mathfrak{su}(3)&\oplus&\mathfrak{u}(1)
\end{array}\end{equation}
\noindent where $r'\in\R,$ and $iQ$ turns out to be \it none other \rm than the electric charge generator,
\begin{equation}iQ:= \frac{1}{6}\left( L_{13}+L_{26}+L_{45}\right)-\frac{1}{2}L_{\epsilon_3}.
\end{equation}

Readers should take note that it is also possible to bypass the $*_{\overrightarrow{\mathbb{H}}}$ step in $\langle 3 \rangle$.  In this case, one may move from the Standard Model's pre-Higgs $\mathfrak{su}(3)\oplus\mathfrak{su}(2)\oplus\mathfrak{u}(1)$ directly to the Standard Model's post-Higgs $\mathfrak{su}(3)\oplus\mathfrak{u}(1)$, simply via $*_{\overrightarrow{\mathbb{C}}}\vspace{1mm}$.  Again, this outcome originally came as a surprise.  

Readers may observe in particular that  \it the familiar complex conjugate is what dictates the Standard Model's final unbroken gauge symmetries.  \rm Conceivably, the Standard Model's gauge group could  have broken in many different ways, had it been paired with different Higgs sectors.  Could the complex conjugate ultimately be stewarding the symmetry breaking process?

\section{Roadmap}

Now that we have charted out certain paths connecting familiar particle models, we will consolidate them  into  one detailed roadmap.  

In this paper, we have listed five different forms of symmetry breaking steps.  Namely, the multivector condition, $\Psi_V$, invariance under two types of octonionic reflection, ${*_{\overrightarrow{\mathbb{O}}}}$ and $*_{\overleftarrow{\mathbb{O}}}$, one quaternionic reflection, $*_{\overrightarrow{\mathbb{H}}},\vspace{.5mm}$ and one complex reflection, $*_{\overrightarrow{\mathbb{C}}}$.  These four types of reflection are precisely those that can be realized in terms of the \it left\rm-multiplication algebra of $\RCHO.$  

The quaternionic reflection $*_{\overleftarrow{\mathbb{H}}},\vspace{.5mm}$ related to right multiplication, is isolated in the sense that, unlike the others, it is not expressible in terms of the left multiplication algebra.  In the second article of this series,~\citep{fr2}, we will see how it plays an important role in the description of spacetime symmetries.

\begin{figure*}[ht]
\begin{center}
\includegraphics[width=12cm]{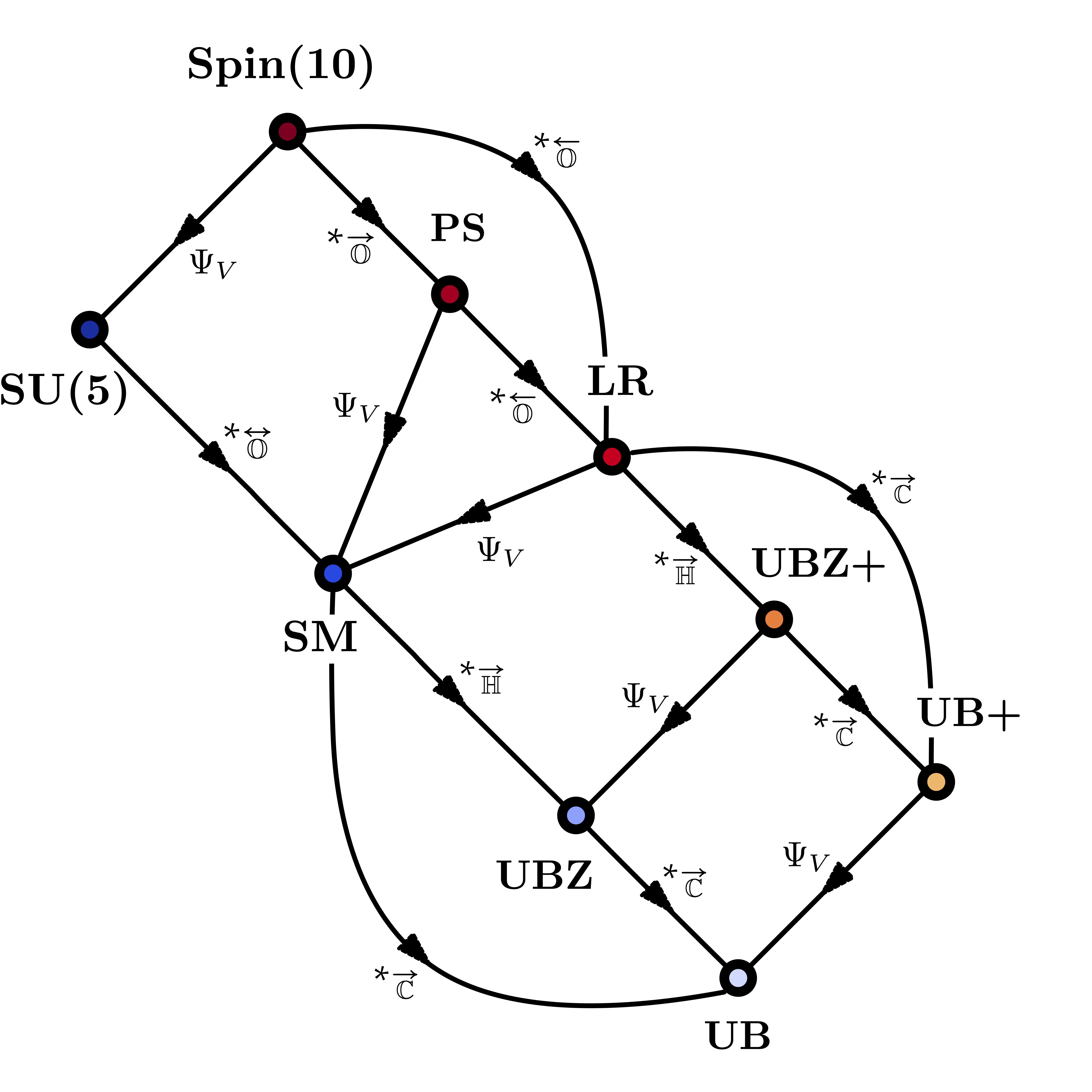}
\caption{\label{5step}  
Six well-known models of elementary particles are interconnected algebraically into a detailed particle roadmap.  The ``SO(10)" model (Spin(10)), the Georgi-Glashow model (SU(5)),  the Pati-Salam model (PS), the Left-Right Symmetric model (LR), the Standard Model pre-Higgs-mechanism (SM), and the Standard Model post-Higgs-mechanism (UB) are each interrelated via certain algebraic constraints.  The \it multivector condition \rm is represented by edges labeled as $\Psi_V$.   \it Division algebraic reflection \rm constraints are represented by edges labeled as $*_{\mathbb{D}}$.  Beyond these six well-known models, there are three more.  For the definition of UBZ+ see equation~(\ref{ubz+}); for the definition of UB+ see equation~(\ref{ub+}); for the definition of UBZ see equation~(\ref{ubz}).}
\end{center}
\end{figure*}

Our culminating diagram results when  all five symmetry breaking steps are amalgamated ($\Psi_V$, ${*_{\overrightarrow{\mathbb{O}}}}$, $*_{\overleftarrow{\mathbb{O}}}$,  $*_{\overrightarrow{\mathbb{H}}},$ $*_{\overrightarrow{\mathbb{C}}}$).  For reflections, we begin with the octonions, continue to the quaternions, and close with the complex numbers.  It is interesting to note that the order $\mathbb{O}\mapsto \mathbb{H} \mapsto \C$ does indeed matter, otherwise one would not necessarily expect our results to hold.  Please see Figure~(\ref{5step}) at the end of this article.  

From the outset, readers may notice an interesting feature of this 5-step  network.  Namely, certain pairs of nodes accommodate multiple pathways between them.  

As a first example, we identify two compatible symmetry breaking pathways from the Spin(10) model to the Left-Right Symmetric model.  One may first restrict Spin(10) to those symmetries invariant under the octonionic reflection $*_{\overrightarrow{\mathbb{O}}}$, resulting in the Pati-Salam model.  Subsequently imposing invariance under ${*_{\overleftarrow{\mathbb{O}}}}$ then results in the Left-Right Symmetric model.  As an alternate route, readers can bypass directly to the Left-Right Symmetric model by simply restricting Spin(10) once via $*_{\overleftarrow{\mathbb{O}}}$, as in~\citep{boyle1}.  

Of more immediate experimental relevance is the bifurcated symmetry breaking step from the Standard Model symmetries pre-Higgs mechanism, $\mathfrak{su}(3)_{\textup{C}}\oplus\mathfrak{su}(2)_{\textup{L}}\oplus\mathfrak{u}(1)_{\textup{Y}}$, to the Standard Model symmetries post-Higgs mechanism, $\mathfrak{su}(3)_{\textup{C}}\oplus\mathfrak{u}(1)_{\textup{Q}}$.   Readers will encounter a direct path via $*_{\overrightarrow{\mathbb{C}}}\vspace{.5mm}$ from $\mathfrak{su}(3)_{\textup{C}}\oplus\mathfrak{su}(2)_{\textup{L}}\oplus\mathfrak{u}(1)_{\textup{Y}}\mapsto\mathfrak{su}(3)_{\textup{C}}\oplus\mathfrak{u}(1)_{\textup{Q}}$.   This algebraic transition coincides with the Standard Model's famous spontaneous symmetry breaking.  On the other hand, 
these same endpoints may be realized by first imposing an intermediate constraint  under $*_{\overrightarrow{\mathbb{H}}}$.  This extra step acts to further break $W^{\pm}$ bosons away from the $Z^0$ relative to the Standard Model.  It will be interesting to investigate whether or not such an intermediate symmetry breaking step could alter the  $W^{\pm}$ mass prediction, relative to that of the Standard Model~\citep{Wmass_nc}, \citep{Wmass_atlas}.

\section{Outlook}

With this framework of particle theories now set in place, one might wonder if there is a common thread connecting each of the symmetry breaking steps displayed here.  Indeed, there is.  

As pointed out in~\citep{it3}, the multivector condition ($\Psi_V$) may be seen to be equivalent to \it fixing a volume element \rm in the exterior/Clifford algebraic representation of $\Psi$.  Likewise, as pointed out in~\citep{fh2}, each of the division algebraic reflection steps $\{ {*_{\overrightarrow{\mathbb{O}}}}, *_{\overleftarrow{\mathbb{O}}},  *_{\overrightarrow{\mathbb{H}}}\}$ may be seen to be equivalent to \it fixing a volume element \rm in the multiplication algebras (associated Clifford algebras) of the various division algebras.  (The complex volume element is trivially preserved.)   The connection of these volume elements to Hodge duality and Jordan algebras will be explored in future work.

\section{Conclusion}
 
In this article, we set out to understand the Standard Model of particle physics in relation to a number of neighbouring particle models.  We summarize our findings:

In what may be viewed as a consistency condition, we introduced what we call the \it multivector condition. \rm This constraint dictates that minimal left ideal fermions transform in accordance with the multivectors that comprise them.  As a result, for internal symmetries, we find that the adjoint, vector, and spinor representations now each transform infinitesimally under the commutator.  The minimal left ideal spinors that we constructed constitute a special case where the commutator and left multiplication coincide.

 The multivector constraint, \citep{abstract}, put an end to the lingering B-L and chirality issues that once haunted~\citep{fh2}.  It also offered one way to understand and consolidate the Spin(10) $\mapsto$ SU(5) symmetry breaking steps explained in~\citep{AGUTS}.
 
The findings of Baez and Huerta, \citep{AGUTS}, and Furey and Hughes, \citep{fh1}, \citep{fh2}, are reconstructed, extended, and amalgamated into a single particle roadmap.  A total of nine particle models are shown to be interlinked via the \it multivector condition, \rm and  via constraints of invariance under generalized \it division algebraic reflections. \rm  Six of these models are familiar to particle physicists, namely the ``SO(10)" model, the Georgi-Glashow model, the Pati-Salam model, the Left-Right Symmetric model, the Standard Model pre-Higgs mechanism, and the Standard Model post-Higgs mechanism.
 
 The algebraic relations between Standard Model pre- and post-Higgs may be of special interest to phenomenologists.  That is, we find it \it fortunate \rm that the complex conjugate singles out precisely those $\mathfrak{su}(3)_{\textup{C}}\oplus \mathfrak{u}(1)_{\textup{Q}}$ gauge symmetries found at low energies, including correct assignments of electric charge, \citep{fr2}.  We find it \it interesting \rm that an alternative symmetry breaking path to this same endpoint may exist, bypassed, and perhaps overshadowed, by the Higgs' direct route. \it  Could such a parallel path carry with it phenomenological implications for electroweak physics? \rm

\begin{acknowledgments}  

These manuscripts have benefitted from numerous discussions with Beth Romano.  The author is furthermore grateful for feedback and encouragement from John Baez, Sukruti Bansal, John Barrett, Latham Boyle,  Hilary Carteret, Mia Hughes, Kaushal Kumar, Agostino Patella, Shadi Tahvildar-Zadeh, Carlos Tamarit, Andreas Trautner, and Jorge Zanelli.

This work was graciously supported by the VW Stiftung Freigeist Fellowship, and Humboldt-Universit\"{a}t zu Berlin.


\end{acknowledgments}

\medskip


\begin{thebibliography}{11}

\bibitem{AGUTS}  Baez, J., Huerta, J., ``The Algebra of Grand Unified Theories," Bull.Am.Math.Soc.47:483-552 (2010) \hspace{3mm} 	arXiv:0904.1556 [hep-th]


\bibitem{fh1} Furey, N., Hughes, M.J., ``One generation of Standard Model Weyl representations as a single copy of $\RCHO$," Phys.Lett.B, 827, (2022) .   This article was widely circulated amongst colleagues on the 16th of February, 2021. Seminar recording available https://pirsa.org/21030013  \hspace{3mm} 	arXiv:2209.13016 [hep-ph]

\bibitem{fh2} Furey, N., Hughes, M.J., ``Division algebraic symmetry breaking," Phys. Lett. B, 831 (2022).  This article was widely circulated amongst colleagues on the 16th of February, 2021.  Seminar recording available https://pirsa.org/21030013  \hspace{3mm}  	arXiv:2210.10126 [hep-ph]

\bibitem{dirac} Consider, for example, the path integral formulation of quantum mechanics, where the integrand in the path integral is given by $e^{iF/\hbar}$.  Only those paths whose $F$ values vary relatively little \it with respect to each other \rm contribute significantly to the integral.  

Dirac, P.A.M., ``The Lagrangian in Quantum Mechanics," St John's College, Cambridge, p. 69-70 (1932)

\bibitem{relativity} In Special Relativity, four vectors do not lend themselves to observer-independent descriptions, but the SO(3,1)-invariant \it inner product between two four-vectors \rm does.



\bibitem{GGquarks} G\"{u}naydin, M.,   G\"{u}rsey, F., ``Quark structure and the octonions," \rm J. Math. Phys.,  14\rm,  No. 11 (1973)

\bibitem{it1} Barducci, A., Buccella, F., Casalbuoni, R., Lusanna, L., Sorace, E., ``Quantized Grassmann variables and unified theories," Phys. Lett., Vol 67B, no 3 (1977)

\bibitem{it2} Casalbuoni, R.,   Gatto, R., ``Unified description of quarks and leptons," \it Phys. Letters, \rm  88B \rm (1979) 306

\bibitem{it3}  Casalbuoni, R., Gatto, R., ``Unified theories for quarks and leptons based on Clifford algebras," Phys.Lett, Vol 90B, no 1,2 (1979).




\bibitem{Greg2001} Trayling, G., Baylis, W.E., ``A geometric basis for the standard-model gauge group," J. Phys. A:  Math Gen 34 (2001) 3009-3324

\bibitem{John2006} Barret, J., ``A Lorentzian version of the non-commutative geometry of the standard model of particle physics," J.Math.Phys. 48 (2007) 012303 \hspace{3mm} arXiv:0608221 [hep-th]

\bibitem{Piotr2008} Zenczykowski, P., ``The Harari-Shupe preon model and nonrelativistic quantum phase space", \it Phys. Lett. B, \rm  660 \rm 567-572 (2008)

\bibitem{Alain2013}  Chamseddine, A.H., Connes, A., van Suijlekom, W.D.,  ``Beyond the spectral standard model: emergence of Pati-Salam unification," JHEP 1311, 132 (2013) \hspace{3mm} arXiv:1304.8050 [hep-th]


\bibitem{Stoica2017}  Stoica,O.C., ``The standard model algebra - leptons, quarks, and gauge from the complex clifford algebra Cl6,''  Adv. Appl. Clifford Algebras (2018) 28: 52 \hspace{3mm}  arXiv:1702.04336 [hep-th]




\bibitem{Brage2020}  Gording, B.,  Schmidt-May, A.,  ``The unified standard model," Advances in Applied Clifford Algebras, (2020) \hspace{3mm} arXiv:1909.05641 

\bibitem{Ivan2020} Todorov, I., ``Superselection of the weak hypercharge and the algebra of the Standard Model,'' JHEP04(2021)164 \hspace{3mm} 	arXiv:2010.15621 [hep-ph]

\bibitem{Ivan2021} Todorov, I.,  ``Clifford algebra of the Standard Model,'' Citeable conference presentation for the Perimeter Institute conference:  Octonions and the Standard Model, PIRSA:21030014  


\bibitem{Norma2023} Bor\v{s}tnik, N.S.M.,  ``How Clifford algebra helps understand second quantized quarks and leptons and corresponding vector and scalar boson fields, opening a new step beyond the standard model," \hspace{3mm} 	arXiv:2306.17167




\bibitem{Conway1937} Conway, A., ``Quaternion treatment of relativistic wave equation,''
Proc. R. Soc. Lond. Ser. A, Math. Phys. Sci., 162 (909) (1937)

\bibitem{Silagadze1994}  Silagadze, Z.K., ``SO(8) Colour as possible origin of generations,'' Phys.Atom.Nucl.58:1430-1434,1995 \hspace{3mm}	arXiv:hep-ph/9411381


\bibitem{Steve1996} Adler, S., ``Quaternionic Quantum Mechanics and Noncommutative Dynamics," \hspace{3mm} arXiv:hep-th/9607008

\bibitem{Stefano1996} De Leo, S.,  ``Quaternions for GUTs," Int.J.Theor.Phys., 35\rm :1821, (1996)

\bibitem{Toppan2003} Carrion, H.L., Rojas, M., Toppan, F., ``Quaternionic and Octonionic Spinors.  A Classification," 2003 (04), 040, \hspace{3mm} arXiv:hep-th/0302113


\bibitem{Baez2011} Baez, J.C., ``Division algebras and quantum theory," Found. Phys. 42 (2012), 819-855 \hspace{3mm} 	arXiv:1101.5690 [quant-ph]

\bibitem{Duff2014} Anastasiou, A.,  Borsten, L.,  Duff, M.J., Hughes, M.J., Nagy, S., ``A magic pyramid of supergravities," JHEP 04 (2014) 178 \hspace{3mm} 	arXiv:1312.6523 [hep-th]

\bibitem{mia} Hughes, M., ``Octonions and Supergravity," PhD Thesis, Imperial College London, 2016. 

\bibitem{Catto2017} Burdik, C.,  Catto S., G\"{u}rcan, Y.,  Khalfan, A.,  Kurt, L., ``Revisiting the role of octonions in hadronic physics,"  Phys.Part.Nucl.Lett. 14 (2017) 2, 390-394

\bibitem{Niels2018} Gresnigt, N., ``Braids, normed division algebras, and Standard Model symmetries," Phys.Lett.B  783 \rm (2018)

\bibitem{Torsten2019} T. Asselmeyer-Maluga, ``Braids, 3-manifolds, elementary particles: Number theory and symmetry in particle physics," Symmetry 11 (10), 1298 (2019) 

\bibitem{Pasha2019} Bolokhov, P., ``Quaternionic wavefunction," IJMPA 34 (2019), 1950001 \hspace{3mm} 	arXiv:1712.04795 [quant-ph]

\bibitem{Tejinder2021} Vaibhav, V.,  Singh, T., ``Left-Right symmetric fermions and sterile neutrinos from complex split biquaternions and bioctonions," Adv. Appl. Clifford Algebras 33, 32 (2023)
\hspace{3mm} 	arXiv:2108.01858 [hep-ph]

\bibitem{boyle1} Boyle, L., ``The Standard model, the exceptional Jordan algebra, and triality," \hspace{3mm}	arXiv:2006.16265

\bibitem{David2021} Jackson, D., ``$\mathbb{O}$ and the Standard Model," Octonions and Standard Model workshop, Perimeter Institute (2021)  https://pirsa.org/21050004

\bibitem{Bruce2022} Hunt, B., ``Exceptional groups and their geometry," in preparation.

\bibitem{Anthony2023}  Lasenby, A., ``Some recent results for SU(3) and octonions within the geometric algebra approach to the fundamental forces of nature,"  Mathematical Methods in the Applied Sciences, (2023)

\bibitem{Corinne2022} Manogue, C.A., Dray, T., Wilson, R.A., ``Octions:  An $E_8$ description of the Standard Model," 	J. Math. Phys. 63, 081703 (2022) \hspace{3mm} 	arXiv:2204.05310 [hep-ph]

\bibitem{ms} J. Schray, C. Manogue, ``Octonionic representations of Clifford algebras and triality", Found.Phys.26:17-70 (1996)  \hspace{3mm} arXiv:hep-th/9407179


\bibitem{Roger2023} Penrose, R., ``Quantized Twistors, G2*, and the Split Octonions," Springer (2022)  

Penrose, R., ``Basic Twistor Theory, Bi-twistors, and Split-octonions," Octonions, Standard Model, and Unification lecture series (2023)  https://youtu.be/xHPfnC9XAjg


\bibitem{Hun2023} Jang, H., ``Gauge Theory on Fiber Bundle of Hypercomplex Algebras,"Nucl. Phys. B 993 (2023) 116281 \hspace{3mm}  arXiv:2303.08159

\bibitem{Basil2023} Hiley, B., ``Dyson's  3-Fold way Quantum Processes and Split Quaternions," Octonions, Standard Model, and Unification, (2023)  https://youtu.be/K5jbsjT6Llk




\bibitem{Dixon1999} Dixon, G., ``(1,9)-spacetime --> (1,3)-spacetime: Reduction => U(1)xSU(2)xSU(3),'' (1999) \hspace{3mm} arXiv:hep-th/9902050


\bibitem{Dixon_recent} Dixon, G.,   ``Division Algebras; Spinors; Idempotents; The Algebraic Structure of Reality'', \rm \hspace{3mm} arXiv:1012.1304v1.

\bibitem{Carlos2019} C. Castro Perelman, ``RCHO-valued gravity as a grand unified field theory," Adv.Appl.Clifford Algebras 29 \rm (2019) no.1, 22 

\bibitem{David2023} Chester, D.,  Marrani, A.,  Corradetti, D.,  Aschheim, R.,  Irwin, K., ``Dixon-Rosenfeld Lines and the Standard Model," \hspace{3mm} arXiv:2303.11334 [hep-th]

\bibitem{Jens2023} K\"{o}plinger, J., ``Towards autotopies of normed composition algebras in algebraic Quantum Field Theory," in preparation.

\bibitem{fr2} Furey, N.,  ``An Algebraic Roadmap of Particle Theories, Part II:  Theoretical Checkpoints," in preparation.


\bibitem{fr3} Furey, N., ``An Algebraic Roadmap of Particle Theories, Part III: Intersections," in preparation.










\bibitem{so10hist} Georgi, H., Zierler, D., AIP Interview 2021, https://www.aip.org/history-programs/niels-bohr-library/oral-histories/44877

\bibitem{su5}  Georgi, H., Glashow, S.L., ``Unity of all elementary particle forces", PRL, vol 32, no 8 (1974)

\bibitem{PS} Pati, J. C., Salam, A.,  ``Lepton number as the fourth ``color"", Physical Review D. 10 (1): 275 289 (1974)





\bibitem{abstract} Furey, N., Romano, B., ``Spinor m-vector constraint:  a new line in the subway of particle models,"  Abstract:  
In his recent talk for this conference, John Baez characterized the Standard Model gauge group as the subgroup of Spin(10) which preserves (1) an $\mathbb{R}^{10}$ splitting into $\mathbb{R}^6\oplus \mathbb{R}^4$, (2) a chosen complex structure, and (3) a complex volume form.  And so we are left with the riddle: could there be a way to characterize the same symmetry breaking patterns from a model based on normed division algebras?  In this talk, we will describe how a new ``spinor m-vector constraint" can provide an alternative to the duo of complex structure and volume form conditions.  Finally, we demonstrate how it is possible to travel from $\mathfrak{so}(10)$ to Pati-Salam (or left-right symmetric) and $\mathfrak{g}_{\textup{SM}}$ to $\mathfrak{su}(3)_{\textup{C}} \oplus \mathfrak{u}(1)_{\textup{Q}}$ by requiring invariance under involutions that generalize the notion of complex conjugation."  

Note that ``m-vector constraint" stands for ``multi-vector constraint".

Abstract circulated widely by organizers Kirill Krasnov and Latham Boyle, scheduled for the 10th of May 2021.
\color{black}






\bibitem{FH} Fulton, W., Harris, J., ``Representation theory, A first course," Graduate texts in Mathematics, Springer (1991) p. 303-307

\bibitem{BM} Burgess, C., Moore, G., ``The Standard Model, a primer", Cambridge University Press (2011)




\bibitem{thesis} Furey, C., ``Standard model physics from an algebra?"  \rm PhD thesis, University of Waterloo, 2015.  \hspace{3mm}www.repository.cam.ac.uk/handle/1810/254719   \hspace{2mm}  	arXiv:1611.09182 [hep-th]

\bibitem{Gen} Furey, C., ``Generations:  three prints, in colour," \rm JHEP,   10\rm, 046  (2014) \hspace{0.3cm} arXiv:1405.4601 [hep-th]

\bibitem{321} Furey, C., ``Three generations, two unbroken gauge symmetries, and one eight-dimensional algebra," \rm Phys.Lett.B, 785 (2018), pp. 84-89  \hspace{3mm}  	arXiv:1910.08395 

\bibitem{malala} Furey, C., ``$SU(3)_C\times SU(2)_L\times U(1)_Y (\times U(1)_X)$ as a symmetry of division algebraic ladder operators," \rm Eur.Phys.J. C,  78 \rm 5 (2018) 375





\bibitem{Wmass_nc} Athron, P., Fowlie, A.,  Lu, C.-T.,  Wu, L., Wu, Y., Zhu, B., ``Hadronic uncertainties versus new physics for the W boson mass and Muon g-2 anomalies," Nature Communications, 14:659 (2023)

\bibitem{Wmass_atlas} The Atlas Collaboration, ``Improved W boson Mass Measurement using $\sqrt{s} = 7$ TeV Proton-Proton Collisions with the ATLAS Detector," ATLAS-CONF-2023-004 







\end{thebibliography}
\end{document}